\documentclass[12pt]{article}
\usepackage[dvips,dvipdf]{graphicx}
\usepackage[usenames]{color}
\usepackage{epsf}
\usepackage{amssymb}
\title{Low energy scattering parameters from the solutions of the non-relativistic 
Yukawa model on a 3-dimensional lattice}

{\author{F. de Soto$^{a,b}$\footnote{fcsotbor@upo.es}, 
J. Carbonell$^a$\footnote{carbonel@lpsc.in2p3.fr}\\
{\small\em $^a$Laboratoire Physique Subatomique et Cosmologie, 
               53 av. des Martyrs, 38026 Grenoble, France}\\
{\small\em $^b$Dpto. Sistemas F\'{\i}sicos, Qu\'{\i}micos y
Naturales, U. Pablo de Olavide, 41013 Sevilla, Spain}
		   }}

\begin{document}
\maketitle
\bibliographystyle{unsrt}

\begin{abstract}
The numerical solutions of the non-relativistic Yukawa model on a 3-dimensional 
size lattice with periodic boundary conditions are obtained.
The possibility to extract the corresponding -- infinite space -- low energy parameters
and bound state binding energies from  eigensates computed at finite lattice size is discussed. 
\end{abstract}


\section{Introduction}\label{intr}

The description  of
deconfined hadronic states from a low energy confining theory is an exciting theoretical problem and
a challenging numerical task. 
With the increasing power of
computers devoted to Lattice QCD and the progress made by
algorithms, it is nowadays possible to study low energy
hadron-hadron systems from first QCD principles.

Contrarily to hadron masses and form factors,  scattering
properties can not be computed from infinite volume euclidean
simulations \cite{MT_NPB_245_90}. 
The link between a field theory formulated in an euclidean space and
the scattering observables is however made possible by studying the
volume dependence of the bound state spectra in a finite box with
periodic boundary conditions.
The underlying formalism was developed by M. Luscher and
collaborators in a series of papers
\cite{ML_CMP104_86,ML_CMP105_86,LW_NPB339_90,ML_NPB354_91,ML_NPB364_91,LL_CMP219_03}
and has been recently updated in view of applications to nuclear
and hadronic physics
\cite{LMST_NPB619_01,BBPS_PLB585_04,BBPS_NPA747_05,BSW_PRD73_06}.

This formalism was formerly applied to extract the
hadron-hadron scattering lengths in the quenched approximation
\cite{FKOMU_PRD52_95}, and more recently  to
obtain the low energy scattering parameters in $\pi\pi$
\cite{CP_PACS_PRD67_03,CP_PACS_PRD70_04,CP_PACS_PRD71_05,BBOS_PRD73_06,BBLOPPS_06},
$K\pi$ \cite{BBLOPPS_06} and even nucleon-nucleon
\cite{BBOS_PRL97_06} systems from fully dynamical lattice QCD calculations. 
These calculations are crucial in providing 
some insight on the fundamental parameters of the hadron-hadron
interaction, as well as low energy observables
which are poorly known  since hardly accessible experimentally.

\bigskip
The aim of our work is to test the applicability of Luscher relations in a model
which, on one hand, would contain the main physical ingredients of hadron-hadron interaction
and, on another hand, would be simple enough to be independently controlled.
This is provided by the Yukawa model, in which 
a massive  scalar particle is exchanged between two fermions. 
When generalized to pseudoscalar and vector exchanges, 
it constitutes the keystone of baryon-baryon interaction models \cite{Nijmegen}. 

In the present paper we have considered the non relativistic reduction
of the Yukawa model that we have solved in a
3-dimensional $L^3$ lattice with periodic boundary conditions.
The solutions have been obtained  using finite difference schemes in close analogy with 
the methods used in lattice field theory simulations.
The  $L$-dependence of the eigenenergies have been used
to extract the infinite volume low energy parameters, namely the scattering length and 
the effective range.
Contrary to the quantum field lattice calculation,
these quantities can be besides accurately computed by solving the corresponding 
one-dimensional Schrodinger radial equation in an independent way.

Despite its simplicity, the non relativistic
Yukawa model considered here contains an essential ingredient of a realistic interaction 
-- its finite range -- which plays a relevant role
in view of extracting the low energy parameters from the finite volume spectra.
It offers a wieldy and physically sound tool to test the validity of the different
approaches discussed in the literature, in particular the large and small $L$-expansion.
The full quantum field contents of this model have also been considered in 
a lattice calculation. Preliminary results can be found in \cite{dSCRBLP_QCD05_05}.

\bigskip
A similar study was undertaken in \cite{BBPS_PLB585_04} in the framework of an effective
field theory without pions 
EFT($\slash\hspace{-2.mm}\pi$)  and from a slightly different point of view.
These authors fix the values of the low energy parameters and obtain the energy levels  
assuming they satisfy Luscher equations \cite{ML_CMP104_86,ML_CMP105_86,ML_NPB354_91}
while in our case all these quantities are independently generated by solving a dynamical model
and used to find their applicability conditions.

\bigskip
The properties of the non relativistic Yukawa model in the continuum are briefly 
introduced in Section 2.
Section 3 is devoted to explain the numerical methods
used for its solution on a 3-d torus. Results concerning 
scattering and bound states are displayed in Section 4
and some final remarks are given in the conclusions.

\section{The model}\label{Model}

\begin{figure}[h!]
\begin{center}
\mbox{\epsfxsize=7.cm\epsffile{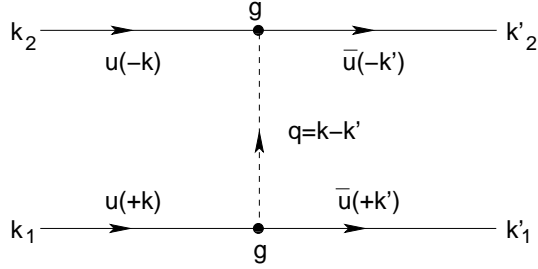}}
\caption{Lowest order two-body amplitude
of the Yukawa model (\ref{Lint}) leading to potential (\ref{V}).}\label{diag_wc}
\end{center}
\end{figure}

We consider a  system of two non relativistic  particles
interacting  by a Yukawa potential of strength $g$ and range parameter $\mu$
\begin{equation}\label{V}
U(r) =- \frac{g^2}{4\pi}\; {e^{-\mu r} \over r}
\end{equation}
This potential is obtained by Fourier transforming the Born amplitude $T_B$
\[ T_B(k_1,k_2,k'_1,k'_2)= 
\bar{u}(\vec{k}')u(\vec{k})\;\frac{g^2}{(k-k')^2-\mu^2}\;\bar{u}(-\vec{k}')u(-\vec{k}) \] 
of  the  meson-fermion interaction lagrangian (See figure \ref{diag_wc}):
\begin{equation}\label{Lint}
{\cal L}(x)=g\bar\Psi(x)\;\Phi(x)\;\Psi(x)   
\end{equation}
where $\Psi$ is a Dirac field  with mass M, $\Phi$ a scalar field with mass $\mu$
and in which 
the non relativistic reduction -- $u=1$ $q^2=-\vec{q}^2$ -- has been applied  \cite{Gross_93,PS_95}.
The same potential results from a model with purely scalar fields.

The reduced S-wave radial Schrodinger equation (in $\hbar=c=1$ units)  reads
\begin{equation}\label{Sch_RR}
u_0"(r) + M\left[ E +  \frac{g^2}{4\pi}\; {e^{-\mu r} \over r} \right]u_0(r)=0 
\end{equation}
and the corresponding tridimensional wavefunction is
\begin{equation}\label{Psi}
\Psi(\vec{r})= {1\over\sqrt{4\pi}} {u_0(r)\over r} 
\end{equation}

The solutions of (\ref{Sch_RR}) depend a priori on the three parameters ($M,\mu,g$) but 
written in terms of the dimensionless variable $x=\mu r$, this equation is equivalent to
\begin{equation}\label{Sch_RRA}
u"(x) + \left[ \epsilon + G {e^{-x}\over x} \right] u(x) =0 
\end{equation}
which depends on a unique dimensionless strength parameter
\[  G(M,\mu,g) = {g^2\over4\pi}\; {M\over\mu} \]

In particular, the bound state energy (E) and scattering length ($a_0$) of the two-body system (\ref{Sch_RR})
are given by the scaling relations
\begin{eqnarray} 
\frac{E}{M} &=& \left(\frac{\mu}{M}\right)^2 \epsilon(G) \label{E}\\
 a_0\mu     &=& A_0(G) \label{a_0}
\end{eqnarray} 
where  $\epsilon(G)$ and $A_0(G)$ are respectively the binding energy
and scattering length of the dimensionlees problem (\ref{Sch_RRA}).

\bigskip
We will focus hereafter on finding the solutions of a unit mass particle in the potential
\begin{equation}\label{Vad}
 V(x)= -G {e^{-x}\over x}
\end{equation}
what we will call the dimensionless Yukawa model.
All length parameters involved must be therefore understood in units of the 
inverse exchanged mass $\mu^{-1}$.
The lagrangian (\ref{Lint}) gives rise to 
an attractive potential, but disregarding its relation with 
the underling field theory, the repulsive case $G<0$ can be considered as well. 

\bigskip
\newcommand\tendto{\mathop{\sim}}

The asymptotic norm $N_s$ of a bound state with energy $\epsilon_0$ is defined as
\begin{equation}\label{N_s}
N_s =\lim_{x\to\infty}  u_0(x) e^{+k_0 x}  \qquad k_0=\sqrt{-\epsilon_0} 
\end{equation}
with $u_0$ normalized by
\[ \int_0^{\infty} u^2_0(x) dx=1\]

The convention used for the scattering length is such that the regular zero-energy solution
of (\ref{Sch_RRA}) is asymptoticaly  given by 
\begin{equation}\label{u0as}
u_0(x) \tendto_{x\rightarrow\infty}  x+A_0 
\end{equation}  
or equivalently the low-energy (S-wave) phase-shifs given by 
\[\delta_0(k)=-A_0k +o(k^2)  \qquad k^2=\epsilon\]
It can be shown  that, in the limit of weak coupling constant,  one has 
\begin{equation}\label{A0G}
 A_0(G) = -G +o(G^2)
\end{equation}  
which correspond to the Born approximation of the Schrodinger equation
with the Yukawa model (see Appendix \ref{Ap1}).

\begin{figure}[h!]
\begin{center}
\includegraphics[width=8.5cm]{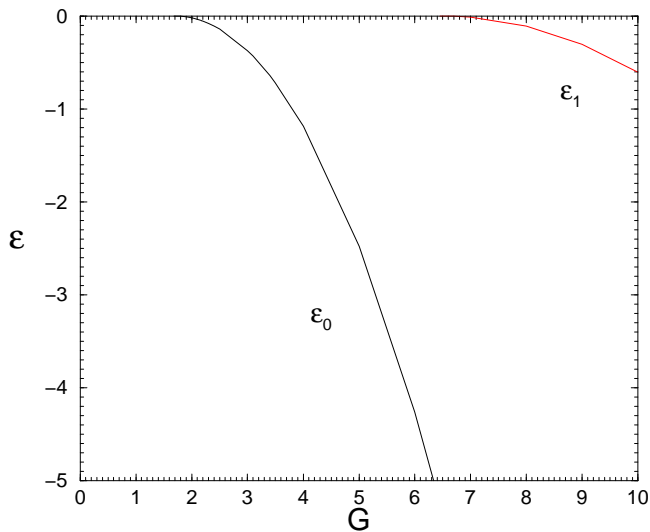}
\caption{Bound state binding energies for the ground ($\epsilon_0$) and first
excited ($\epsilon_1$) S-wave state as a function of
the coupling constant $G$ in the dimensionless Yukawa model (\ref{Vad})}
\label{eps_G}
\end{center}
\end{figure}

\begin{figure}[h!]
\begin{center}
\includegraphics[width=10.cm]{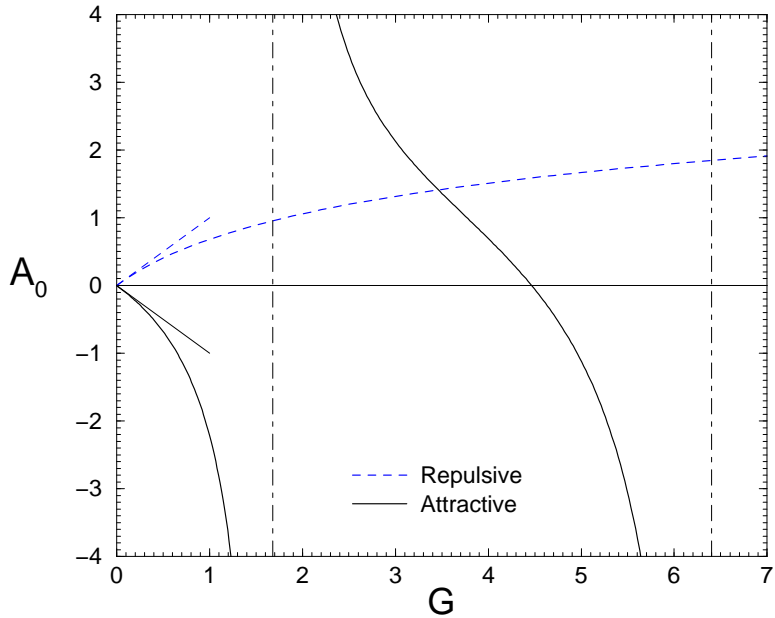}
\caption{Scattering length $A_0$ in the dimensionless Yukawa potential (\ref{Vad})
as a function of the coupling constant $G$. 
Dashed line correspond to the repulsive case ($G<0$).
The tangents at the origin indicates the corresponding Born approximation $A_0^{(B)}(G)=\pm G$ 
and the vertical asymptote the two first bound state singularities at $G_0=1.68$
and $G_1=6.44$.}\label{A_G}
\end{center}
\end{figure}

The "universal" functions $\epsilon(G)$ and $A_0(G)$ 
are displayed in figures \ref{eps_G} and  \ref{A_G}.
The system has a first bound state ($\epsilon_0$) for coupling constant $G$ above some critical
value $G_0\approx1.680$ at which $A_0(G)$ has a pole.
An infinity of similar branches -- and corresponding $A_0(G)$ poles --  appear for the excited states: $\epsilon_1$ 
at $G_1=6.44$, $\epsilon_2$  at $G_2=14.34$ \ldots

\bigskip
It will be of further interest to consider also the effective range parameter $R_0$
which appears in the low energy expansion 
\begin{equation}\label{ER}
 k\cot\delta_0(k)=-{1\over A_0} + {1\over2} R_0k^2 + o(k^4)
\end{equation}

It is worth noticing that the solutions obtained by inserting potential (\ref{V}) in the 
Schrodinger equation correspond to summing up the iterations of the diagram 
displayed in figure \ref{diag_wc}, i.e. the so called  "ladder" approximation. 
This represents a small --  though infinite -- part of the diagrams included in the 
interaction lagrangian (\ref{Lint}).
Attemps to include all the  interaction terms of this lagrangian
in two-body bound and scattering states  have been presented in \cite{dSCRBLP_QCD05_05}
in the framework of lattice quantum field calculation.

\bigskip
Though the solutions of (\ref{Sch_RRA}) can be easily obtained with great accuracy,
we are interested on the results provided by the same
numerical schemes than the ones used on lattice calculations.
Thus, by using finite difference methods on an equidistant grid with stepsize $a$, i.e.
\[ a^2\; u"(x)= u(x+a)-2u(x)+u(x-a) \]
one obtains the ground state binding energies $\epsilon_0$ and low-energy scattering parameters  given 
respectively in Tables \ref{Tab_BS} and \ref{Tab_LEP}.

\begin{table}[h]
\begin{center}
\begin{tabular}{l| c | c | c |cc|}
    & a=0.50     & a=0.20	&a=0.10	    &\multicolumn{2}{|c|}{Exact}\\\hline
G   &$\epsilon_0$&$\epsilon_0$&$\epsilon_0$   &$\epsilon_0$&$\rho$ \\\hline
2.0 & -0.0076    & -0.0177	&-0.0198	    & -0.0206    & 5.63\\
2.5 & -0.0791    & -0.1270	&-0.1364	    & -0.1397    & 2.50\\
3.0 & -0.2210    & -0.3386	&-0.3625	    & -0.3710    & 1.69\\
4.0 & -0.6831    & -1.0665	&-1.1531	    & -1.1849    & 1.07\\
\end{tabular}
\caption{Binding energy $\epsilon$ and radius $\rho=\sqrt{<x^2>}$ for the ground state
of the Yukawa model using finite difference methods with stepsize $a$.}\label{Tab_BS}
\end{center}
\end{table}
\begin{table}[h]
\begin{center}
\begin{tabular}{l| ll | ll | ll |ll |}
    &\multicolumn{2}{|c|}{a=0.50} & \multicolumn{2}{|c|}{a=0.20} &\multicolumn{2}{|c|}{a=0.10}&\multicolumn{2}{|c|}{Exact}\\\hline
G   &   $A_0$	 &   $R_0$      &   $A_0$	&   $R_0$	     &    $A_0$	 &   $R_0$      &   $A_0$  & $R_0$  \\\hline
0.10&  -0.1029	 &  40.25	    & -0.1049	& 40.16	     & -0.1052	 &  40.15	    &  -0.1053 & 40.15  \\
0.20&  -0.2170	 &  20.24	    & -0.2227	& 20.14	     & -0.2224	 &  20.13	    &  -0.2226 & 20.12 \\
0.40&  -0.4881     &  10.20       & -0.5015     & 10.10          & -0.5035     &  10.09       &  -0.5041 & 10.08  \\
0.50&  -0.6518	 &  8.180	    & -0.6721	& 8.078	     & -0.6751	 &  8.063	    &  -0.6761 & 8.058 \\
1.00&  -2.034	 &  4.076	    & -2.177	& 3.958	     & -2.199	 &  3.940	    &  -2.207  & 3.934  \\
1.50&  -8.039	 &  2.628	    & -10.81	& 2.488	     & -11.40	 &  2.467	    &  -11.61  & 2.460 \\
\end{tabular}
\caption{S-wave low energy scattering parameters (scattering length $A_0$ and effective range $R_0$
for different lattice step values $a$.}\label{Tab_LEP}
\end{center}
\end{table}

One can remark the different sensibility of the bound and scattering results
with respect to the stepsize. While low energy
parameters vary only of few percent for $a\in[0.1,0.5]$, except near the resonant value $G=1.5$,
the binding energy varies by a more than a factor 2 in the same interval.

\section{Solutions on a torus}

Let us  consider  now the solutions of the dimensionless Yukawa model (\ref{Vad})
on a the 3-d torus, i.e.  the solutions of
\begin{equation}\label{Sch_3D}
\left[ -\Delta_{\vec{x}}  + V_L(\vec{x}) \right] \Phi(\vec{x})=\epsilon(L) \;\Phi(\vec{x})
\end{equation}
satisfying periodic boundary conditions 
\[ \Phi(\vec{x}+\vec{n}aL)=\Phi(\vec{x}) \qquad \forall\vec{n}\in Z^3 \]
where $a>0$ is the lattice step and $L\in N$ is the number of lattice points on each spatial dimension.

The "lattice"  potential $V_L$ is defined as
\begin{equation}\label{VL}
V_L(\vec{x})=\sum_{\vec{n}\in Z^3} V( \vec{x}+\vec{n}aL) 
\end{equation}
where $V$ is the infinite volume interaction given in (\ref{Vad}).
$V_L$ incorporates the same periodicity than the solutions
and contains the interaction with the "surrounding world".

\begin{figure}[h!]
\begin{center}
\includegraphics[width=7.cm]{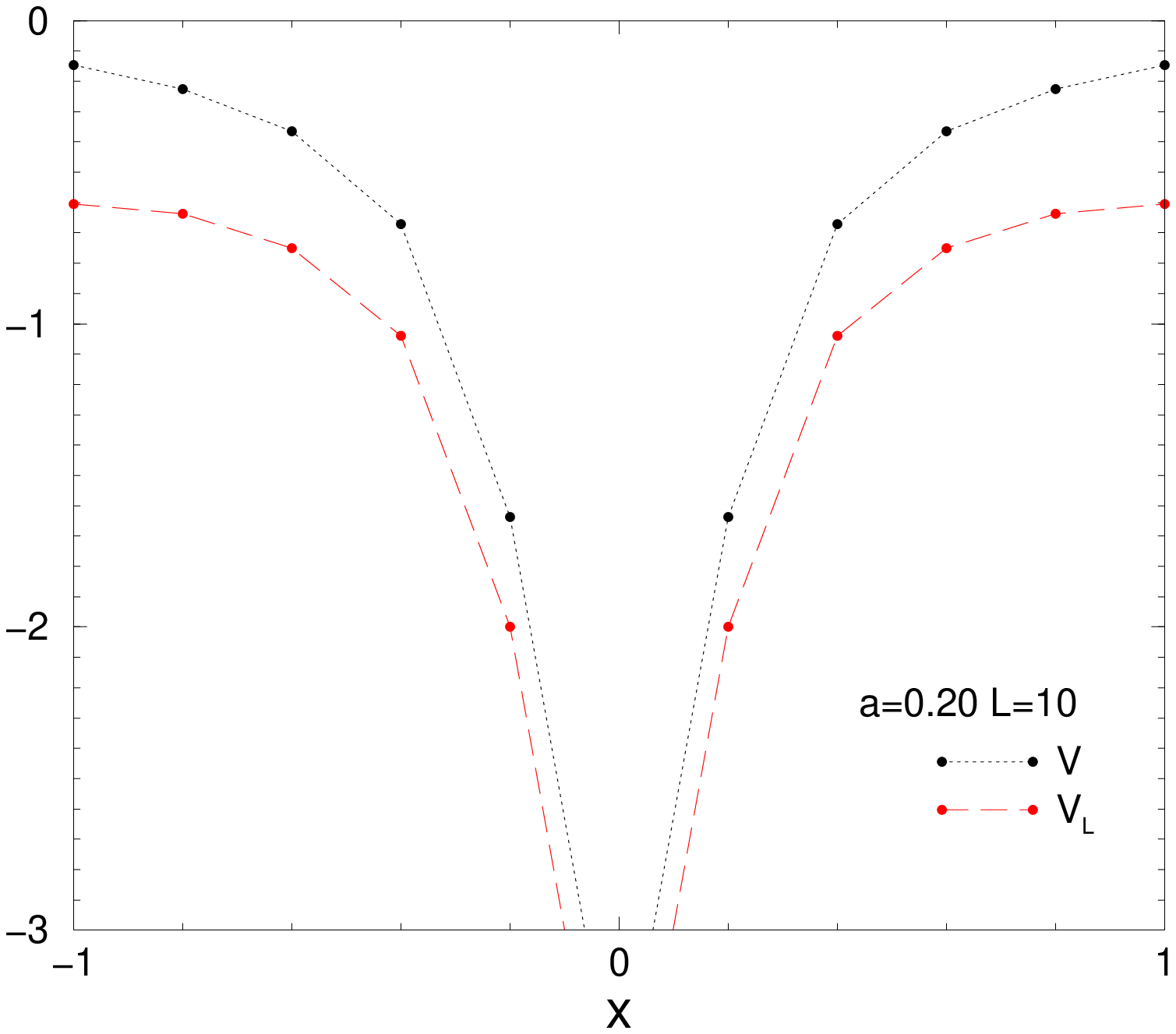}\hspace{1.cm}
\includegraphics[width=7.cm]{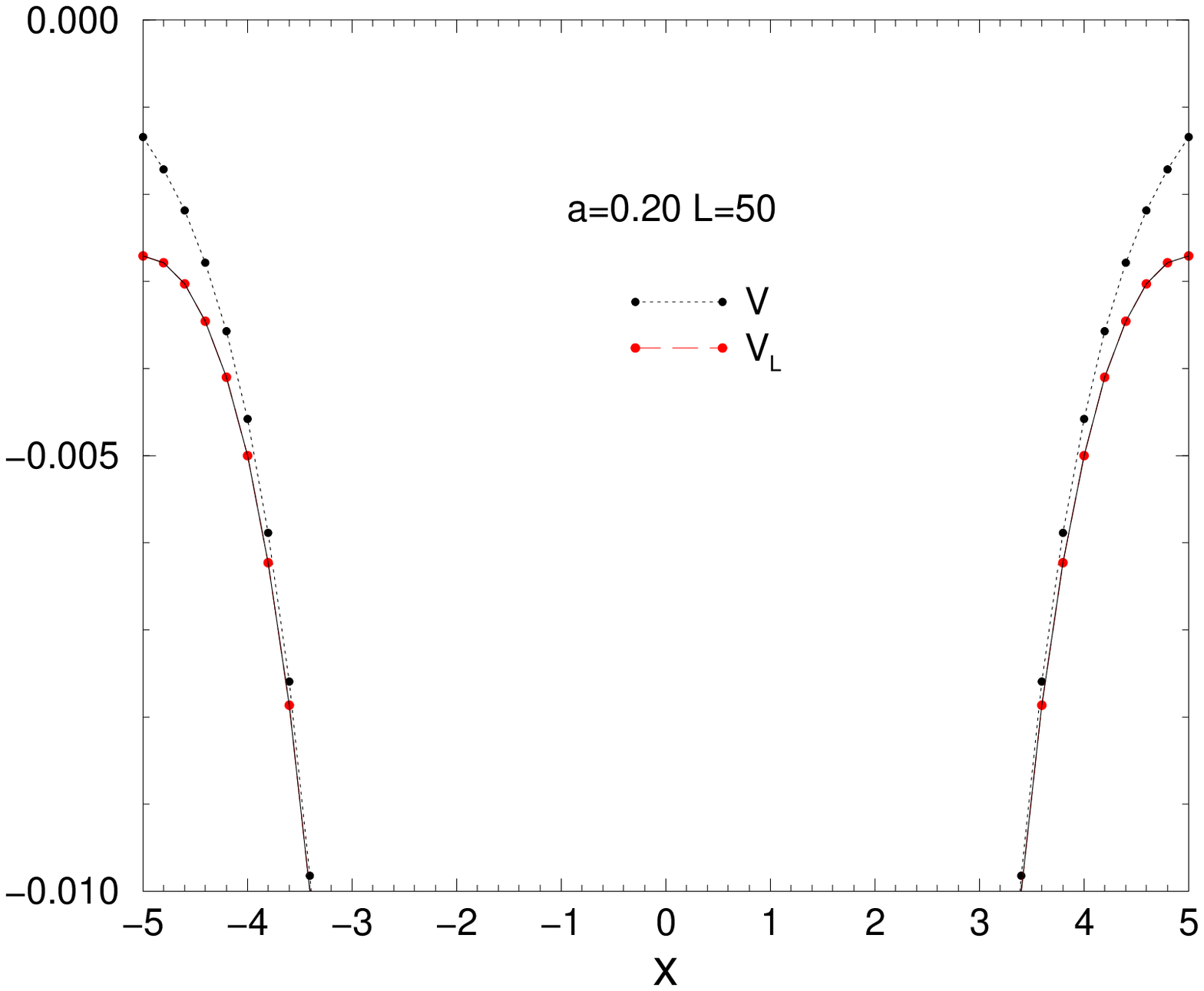}
\caption{The lattice potential $V_L$ defined in (\ref{VL})
is compared with the infinite volume interaction $V$ for $aL=2$ and $aL=10$ }\label{Fig_VL}
\end{center}
\end{figure}

For large values of $aL$, $V_L$ is well approximated by $V$ but below
$aL\sim 10$ the "mirror" contributions are sizeable
and dramatically modify the low-$L$ behaviour of the observables.
We display in figure \ref{Fig_VL} the comparison between $V_L$ 
-- converged with respect to the sum  over mirror images (\ref{VL}) -- and $V$
obtained with $a=0.2$ and different values of $aL$.

\bigskip
In order to obtain the numerical solutions of (\ref{Sch_3D}), we 
introduce in the 3-d cubic lattice the coordinates
\[x_{\mu}(i_{\mu})=a\left( i_{\mu}-{L\over2} \right) \qquad \mu=1,2,3 \quad i_{\mu}=0,1,\ldots,L-1\]
Using the finite difference scheme on gets the following expression for the 3-d laplacian
\begin{equation}\label{Lap}
\Delta\Phi_i = -{6\over a^2}\left\{\Phi_{i}-{1\over6}\sum_{\mu=1,3}\Phi_{i+\mu}+\Phi_{i-\mu}\right\}
\end{equation}
where $i$ labels the set of coordinate indices $i\equiv\{i_1,i_2,i_3\}$
and $i+\mu$ denotes the nearest neighbour of $i$ in the direction $\mu$.

\bigskip
The stationary states can be obtained by solving directly the eigenvalue equation (\ref{Sch_3D}) 
together with (\ref{Lap}).
However, in order to follow as close as possible the
lattice methods we alternatively use an euclidean time-dependent approach.
To this aim we consider the evolution operator between $t$ and $t+\Delta$
\[ U(t,t+\Delta )=e^{-iH\Delta } \]
The time-dependent wave function propagates in this interval according to
\[\Psi(x,t+\Delta)=U(t,t+\Delta)\Psi(x,t)
                    = e^{-iH{\Delta\over2}}e^{-iH{\Delta\over2}}\Psi(x,t)\]
which can be written in the form
\begin{equation}
e^{+{i\Delta\over2}H}\Psi(x,t+\Delta)=e^{-i{\Delta\over2}H}\Psi(x,t)
\end{equation}
For short $\Delta$ values, 
we approximate $U$ by its Taylor expansion up to $\Delta^2$ terms, 
and get the following relation between the wave functions  
at two consecutifs time steps $t$ and  $t+1$
\begin{equation}\label{EV1}
\Psi^{t+1}_j = \frac{1-{i\Delta\over2}H}{1+{i\Delta\over2}H} \; \Psi^t_j 
\end{equation}
where we use the notation $\Psi_j^t\equiv\Psi(x_{\mu}(j),t\Delta)$.

This approximate numerical scheme has the advantage of preserving the unitarity.
Using relation (\ref{EV1}) would however imply the inversion of $1+iH$ which is an unpleasant
task. This can be
avoided by noting that (\ref{EV1}) is actually equivalent  to the system of equations
\begin{eqnarray*}
\left( 1+{i\Delta\over2}H\right) \chi^t_j  &=& 2\Psi^{t}_j  \cr
\Psi^{t+1}_j &=& -\Psi^t_j + \chi^t_j 
\end{eqnarray*}
The unknown  $\chi^t_j$  is a solution of an inhomogeneous Schrodinger-like equation
which can be solved using the same discretization schemes described above for the
stationary states.

\newcommand\approxtendto{\mathop{\approx}}

The binding energies can be then obtained
by propagating an arbitrary initial solution $\eta_j$ in the euclidean time $\tau=it$.
Indeed, by  expanding the initial state in terms of stationary eigenfunctions $\Phi^{(n)}$
\[ \eta =\sum_n c_n \Phi^{(n)}_j\]
it follows that
\[\Psi^{\tau}_j=\sum_n c_n \Phi^{(n)}_j e^{-\epsilon_n\tau\Delta} 
\approxtendto_{\tau\to\infty}c_0 \Phi^{(0)}_j e^{-\epsilon_0\tau\Delta} \]
where $\epsilon_0$ is the ground state energy and $\Phi^{(0)}_j$ the corresponding eigenfunction.
The stationary wavefunction will be normalized according to
\begin{equation}\label{NORM}
(aL)^3\sum_i\Phi^2_i =1
\end{equation}

\begin{figure}[h!]
\begin{center}
\mbox{\epsfxsize=9.cm\epsffile{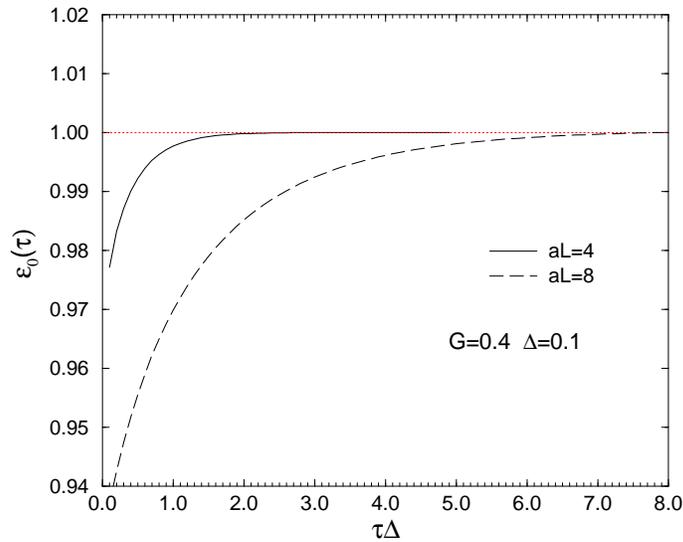}}
\caption{Convergence of the ground state effective energy $\epsilon_0(\tau)$ defined in (\ref{Eef})
as a function of the euclidean time $\tau$ for two different spatial lattice sizes $aL$.
The values are normalized to $\epsilon_0(\infty)$.
The ground state energy is well separated from excited states at $\tau\Delta\approx aL$. 
Results correspond to $G=0.4$ $a=0.2$ and $\Delta=0.1$.}\label{Fig_Eef_t}
\end{center}
\end{figure}

One can get rid of the spatial degrees of freedom by defining at each time-step $\tau$
\[ C(\tau)= {1\over (aL)^3}\sum_j \Psi^{\tau}_j  \sim e^{-\epsilon_0\tau\Delta} \]
This procedure is very similar to the time-slice approach in lattice calculations
and can be used to disentangle the ground state from the first excitation energies.
To this aim it is interesting to define the effective energy of the state
\begin{equation}\label{Eef}
 \epsilon_0(\tau) = {1\over\Delta}\log \frac{C(\tau)}{C(\tau+1)}
 \end{equation}
and study the convergence $\epsilon_0(\tau)\to\epsilon_0$.
The density of states increases
with the size of the box, making their separation increasingly difficult. 
However until moderate lattice sizes $aL\sim 10$ --
values we will be further interested in  -- the energy levels are 
well isolated, what makes easy the convergence of the euclidean-time propagation
of the ground state.
This is illustrated in Figure \ref{Fig_Eef_t}, showing the $\tau$-dependence of the 
ground state effective energy  (\ref{Eef}) -- normalized to $\epsilon_0(\infty)$ -- for two different
$aL$ values  obtained with $G=0.4$, $a=0.2$ and $\Delta=0.1$.
In these cases the results are converged for $\tau\Delta\approx aL$.

\section{Results}

\begin{figure}[h!]
\begin{center}
\mbox{\epsfxsize=7.cm\epsffile{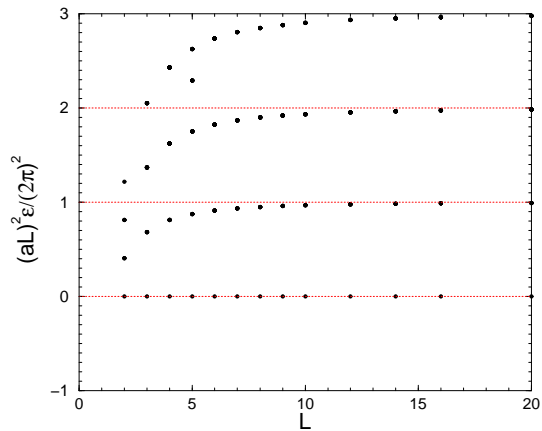}}
\caption{L-dependence of the first free two-body energy states}\label{L2EPSL_G_0}
\end{center}
\end{figure}

Before presenting the results of the Yukawa model for bound and scattering states
we will summarize the free case.
The  non relativistic energies of two -- unit mass -- particle states in a lattice with periodic boundary
conditions are given by
\begin{equation}\label{eps0L}
\epsilon^{(0)}_{n}(L) =  {2\over a^2} \sum_{\mu=1}^3 \left\{ 1-\cos{{2\pi\over L}n_{\mu}} \right\}   
= {4\over a^2} \sum_{\mu=1}^3  \sin^2\left\{{\pi\over L}n_{\mu} \right\}
\end{equation}
with $n_{\mu}=0,1,2\ldots,L-1$.
The continuum limit of an energy state 
$\{n_{\mu}\}$ is reached -- independently of $a$ -- for  $L>>\pi {\rm Max}\{n_{\mu}\}$  with
\[ \epsilon^{(0)}_{n}(L) \approx  p^2 \qquad\vec{p}= {2\pi\over aL} \vec{n}\]

The ground state energy ($n_{\mu}=0$) is zero for any value
of L and the corresponding wavefunction is a constant, which normalized according to (\ref{NORM}), equals
\begin{equation}\label{FGSWF}
\Phi^{(0)}_i={1\over \sqrt{(aL)^3}}
\end{equation}
All excited state energies depend on the lattice size and decreases asymptotically like $1/(aL)^2$, 
a regime already reached with a number of lattice points $L\sim10$ for the first excitation.
The lowest part of the free spectrum is displayed in figure \ref{L2EPSL_G_0}. 
One can remark the high degeneracy of the states.

\begin{figure}[h!]
\begin{center}
\mbox{\epsfxsize=14.cm\epsffile{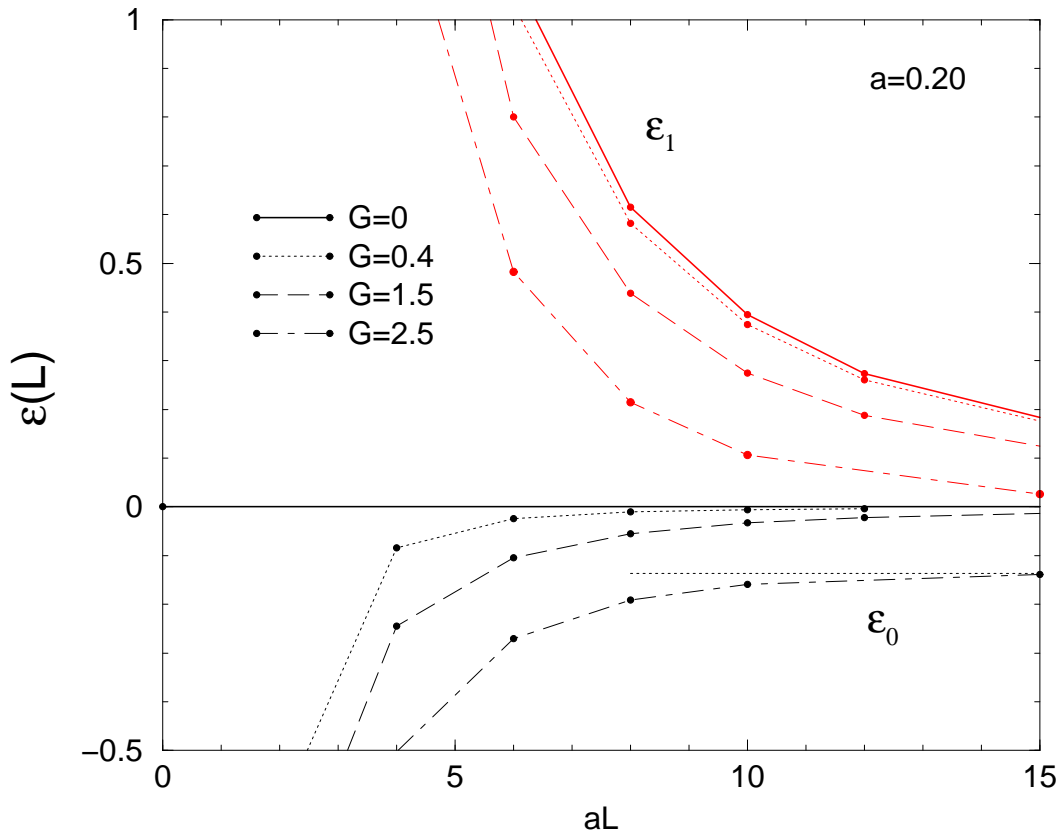}}
\caption{Evolution of the first free $\epsilon_n(L)$ trajectories (solid lines)
with the interaction strength in the Yukawa model. 
Results correspond to $G=0.4$ (dotted) and $G=1.5$ (dashed), both below 
the bound state threshold and $G=2.5$ (dotted-dashed)
where the two-body systems has a bound state with $\epsilon_0=-0.12$.}\label{L2EPSL_G=0.0_0.40}
\end{center}
\end{figure}

The generic evolution of the  free trajectories when increasing the interaction
is displayed in Figure  \ref{L2EPSL_G=0.0_0.40}.
The ground $\epsilon_0(L)$ and first excited  $\epsilon_1(L)$ state energies 
are  plotted as a function of the lattice size $L$ for different values
of coupling constant $G$.

In the attractive case we are considering, the ground state energy $\epsilon_0$
becomes negative and $L$-dependent.
For $G<G_0$, i.e. before the appearance of the first bound
state, the excited states remain with positive energy
and all trajectories tend to zero in the limit $(aL)\to\infty$.
This behaviour is illustrated  
for $G=0.4$ (dotted line) and $G=1.5$ (dashed line) in figure \ref{L2EPSL_G=0.0_0.40}.
In this range all states tend to the free solutions $\epsilon_n(L)\to\epsilon^{(0)}_n(L)$ for large 
enough values of the lattice size.

When $G_0<G<G_1$ the ground state trajectory $\epsilon_0(L)$ tends to a non zero 
negative value
corresponding to the infinite volume binding energy $\epsilon_0$
while the first excited state $\epsilon_1(L)$ tends to zero as for the $G<G_0$ case.
Results corresponding to $G=2.5$ are displayed in figure \ref{L2EPSL_G=0.0_0.40}
in dot-dashed line. 
The sign of the  $\epsilon_1(L)$ will in fact depend on
the value of $G$. For $G$ close to the first bound state pole $G_0$ (see figure \ref{A_G}),
$\epsilon_1(L)$ is always positive while for larger values of $G$, close
to the second bound state singularity $G_1$, $\epsilon_1(L)$  is negative.
The change of sign depends actually on $L$ but for large
enough lattice sizes it corresponds to the zero of $A_0(G)$.
  
\begin{figure}[h!]
\begin{center}
\mbox{\epsfxsize=12.cm\epsffile{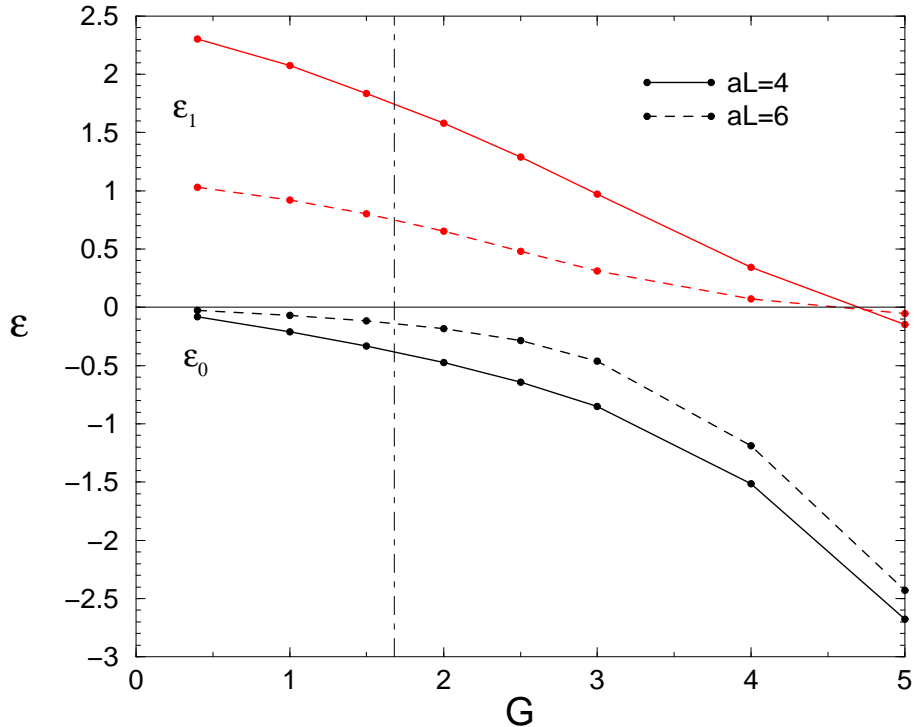}}
\caption{Dependence of $\epsilon_n(L)$ on the coupling constant
at fixed lattice size. 
Results correspond to $aL=4$ (solid line) and $aL=6$ (dashed) with $a=0.1$.}\label{EPSG_L}
\end{center}
\end{figure}

It is interesting to remark that despite  the singular behaviour 
of $A_0(G)$ displayed in figure \ref{A_G}, the $G$-dependence 
of $\epsilon_n(L)$ at a fixed lattice size remains smooth
even when crossing the bound state singularities.
This is illustrated in figure \ref{EPSG_L}
where the $\epsilon_n(G)$ dependence is shown for two different values of $aL$.
Notice the change of sign of $\epsilon_1(L)$ at $G\approx 4.5$
for which $A_0(G)\approx0$.

A striking difference between the free and the interacting case -- not
clearly manifested in figure  \ref{L2EPSL_G=0.0_0.40} --
is the $\epsilon_n(L)$ dependence in the limit of small lattice sizes.
We will see in the next section that
for interacting systems, this dependence is $\sim{G\over (aL)^3}$ 
while for the free case they behave like $\sim{n^2\over (aL)^2}$.

The behaviour described above is qualitatively the same in all the intervals $G_n<G<G_{n+1}$ 
of the coupling constant  for the corresponding $\epsilon_{n}(L)$ state.

Whatever the value of the interaction strength, 
the two-body spectrum in a finite box is always discrete.
The main achievement of Luscher and collaborators is to take profit of the $L$-dependence of $\epsilon_n$ -- 
or more precisely of the differences $\epsilon_n-\epsilon^{(0)}_n$  -- 
to extract the infinite volume low energy parameters and bound state energies. 
Such a possibility  will be discussed in the next subsections.

\subsection{Low energy scattering parameters}

For a coupling constant $G<G_0=1.68$, the two-body system in the infinite volume 
has only scattering states while in a finite box
the spectrum is constituted by a series of discrete values.

In his first work devoted to this subject \cite{ML_CMP105_86}, 
Luscher established a relation between the
two-body  binding energy $\epsilon_n(L)$  on a 3-dimensional spacial lattice with
periodic boundary conditions and the corresponding  scattering length. 
For the S-wave ground state energy $\epsilon_0(L)$ it reads \cite{Note1}
\begin{equation}\label{Luscher_A}
\epsilon_0(L)={4\pi A_0\over(aL)^3} 
\left\{1+c_1\left({A_0\over aL}\right)+c_2\left({A_0\over aL}\right)^2+\ldots\right\}
\end{equation} 
where $A_0$ is the infinite volume scattering length and $c_1=2.837 297$ and $c_2=6.375183$ are universal constants, independent of the details of the particular dynamics.
This relation was proved to be valid in non relativistic quantum
mechanics as well as in quantum field
theory and must be considered as an asymptotic series on powers of $A_0/(aL)$.
For attractive potentials, and with our convention for the scattering length, one has  
$A_0(G<G_0)<0$ and consequently $\epsilon_0(L)<0$.

\bigskip
Equation (\ref{Luscher_A}) is the most popular
of the Luscher relations and has been widely used in lattice calculations 
to extract the value of $A_0$ from a fit to the computed $\epsilon_0(L)$.
We adopt a slightly different point of view by constructing from the 
$\epsilon_0(L)$ values, quantities tending to $A_0$ -- the quantity we are interested in -- for large values of $aL$.
This merely consists in inverting (\ref{Luscher_A}).

To this aim it is interesting to consider slowly varying functions
and to use -- rather than  $\epsilon_0(L)$ -- the combination
\begin{equation}\label{A00}
A_0^{(0)}(L)\equiv {1\over4\pi}\; (aL)^3\epsilon_0(L)
\end{equation}
It  tends asymptotically ($aL\to\infty$)  to
the infinite volume scattering length $A_0$ and
constitutes the zero-th order approximation of
Luscher expansion which can be written as
\begin{equation}\label{Luscher_A00}
A_0^{(0)}(L)=A_0\;\left\{1+c_1\left({A_0\over aL}\right)+c_2\left({A_0\over aL}\right)^2 +\ldots\right\}  
\end{equation}

It is possible to get a series $A^{(n)}_0(L)$ of improved values converging towards $A_0$ by solving equation
(\ref{Luscher_A00}) truncated at the order $n$ for a fixed value of $L$.
One thus obtains, for instance
\begin{equation}\label{Luscher_A03}
A_0^{(2)}(L)=aL  \;z^{(2)} 
\end{equation}
where $z^{(2)}$ is a solution of the cubic equation
\[ z(c_2z^2 + c_1z+1)= {A_0^{(0)}(L)\over aL} \]
 
This expansion is however of small practical interest for it requires lattice sizes
one or two order of magnitude larger than the scattering length, as it was already noticed
in reference \cite{BBPS_PLB585_04}.

\begin{figure}[h]
\begin{minipage}[h!]{8.cm}
\begin{center}\mbox{\epsfxsize=8.cm\epsffile{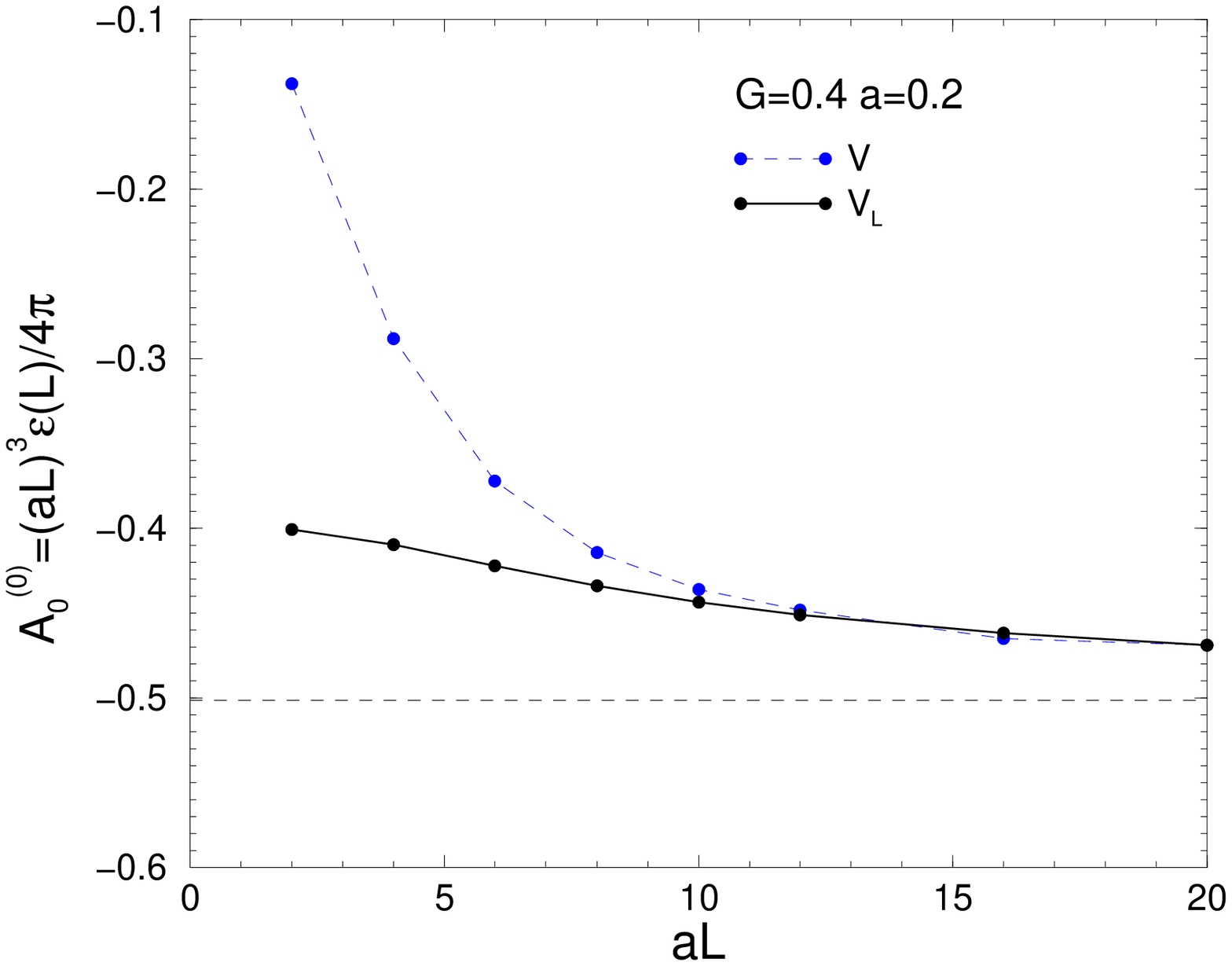}}\end{center}
\caption{Zero-th order scattering length $A^{(0)}_0={(aL)^3\epsilon_0/4\pi}$ for $G=0.4$ and $a=0.2$ obtained
with the infinite volume potential $V$ alone (dashed line) and (solid line) with the full lattice potential $V_L$  (\ref{VL}) }\label{A00L_G=-0.4_LAMBDA}
\end{minipage}
\hspace{0.5cm}
\begin{minipage}[h!]{8.cm}
\begin{center}
\mbox{\epsfxsize=8.cm\epsffile{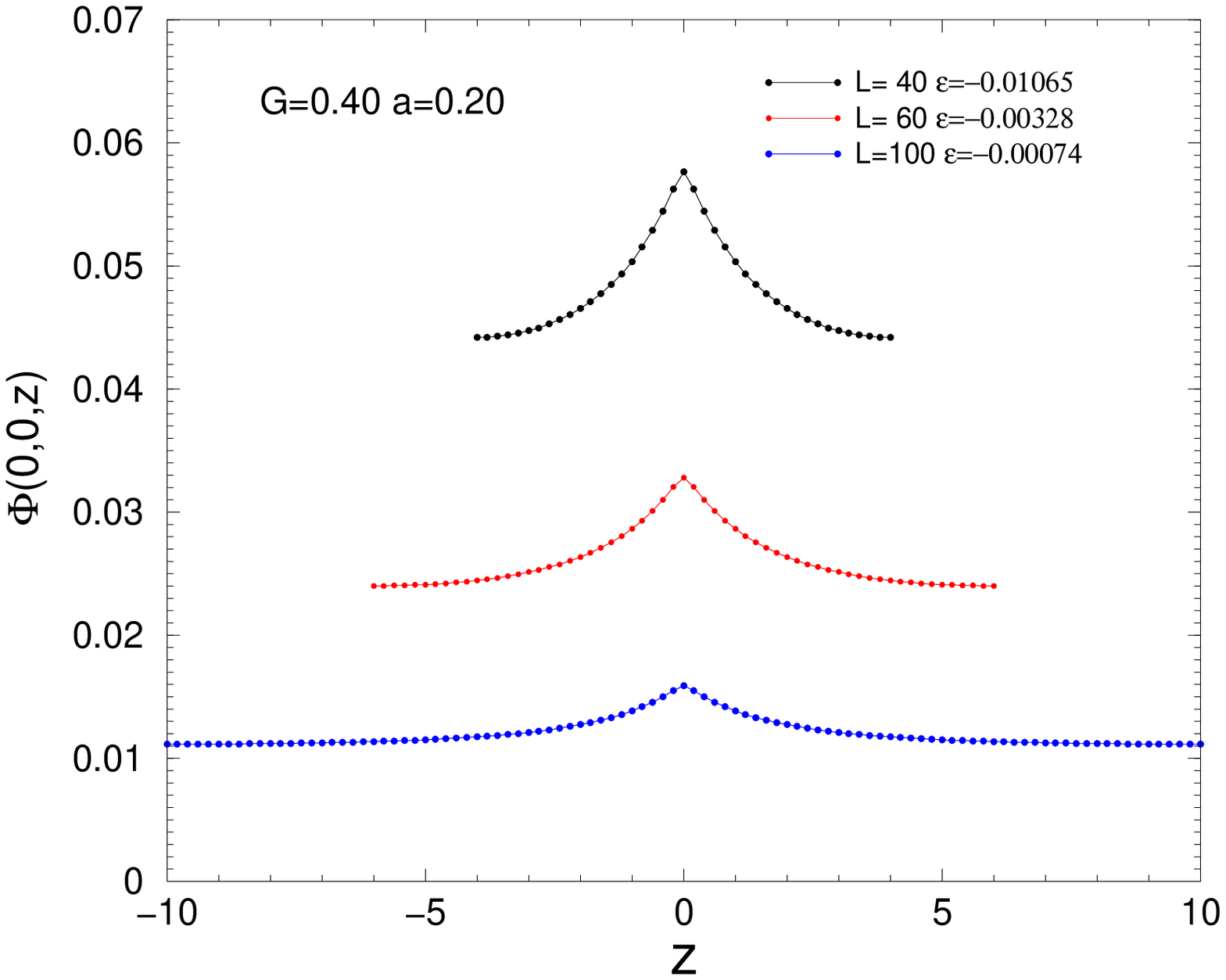}}\end{center}
\caption{Wavefunctions corresponding to 
the results of figure \ref{A00L_G=-0.4_LAMBDA} for selected values
 of L and normalized according to (\ref{NORM})}\label{Phix_G=-0.4_a=0.20}
\end{minipage}
\end{figure}

Figure \ref{A00L_G=-0.4_LAMBDA} represents the $A_0^{(0)}(L)$ results for the ground state 
obtained with $G=0.40$ and $a=0.20$. 
Dashed lines correspond to the solutions with the infinite volume interaction $V$ 
alone and continuous line with the full interaction $V_L$ in (\ref{VL}). 
In both cases, $A^{(0)}_0$ tends indeed asymptotically towards the physical value $A_0=0.501$ 
(horizontal dotted line) in very good agreement with the results
given in table \ref{Tab_LEP} with the same value of $a$,
but the convergence is very slow. Even for computing such a scattering length value, small
with respect to the lattice sizes, a consequent number of grid points would be required.
With $aL=10$ only a 20\% accuracy is reached. 
One can also remark the sizeable effects of the "interactions with the surrounding world"
below $aL\approx10$; these contributions -- which are absent in the pionless EFT considered in \cite{BBPS_PLB585_04} -- are essential 
in providing  the very smooth variation of $A_0^{(0)}(L)$ observed in  the whole range of $aL$.

Corresponding wavefunctions --  normalized according
to (\ref{NORM}) -- for $L=40, 60, 100$ are displayed in Figure \ref{Phix_G=-0.4_a=0.20}.
For small values of $L$
they look similar to the bound states ones (see next subsection) but
their central values depends strongly on the lattice size.
Outside the interaction region the wave function behaves -- according to (\ref{u0as}) --
like $\Phi(x)\sim 1+{A_0\over x}$ 
and would thus provide an alternative way to extract $A_0$ in case it was numerically accessible. 
In the infinite volume limit they spread 
over all the lattice with constant amplitude, like in the free case.

\bigskip
The behaviour illustrated in figure \ref{A00L_G=-0.4_LAMBDA} is in fact generic and easy to understand.
On one hand we have shown (see Appendix \ref{Ap2}) that, under reasonable assumptions, the $aL\to0$ limit of $A_0^{(0)}$ is finite. For the Yukawa model it gives
\begin{equation}\label{A000} 
\lim_{aL\to0} A_0^{(0)}(L)= -G 
\end{equation}

On the other hand, the large $aL$ limit is, by construction,
\[ \lim_{aL\to\infty} A_0^{(0)}(L)=A_0(G)\]
with $A_0(G)$ given in figure \ref{A_G}.
For small values of the coupling constant, the 
L-dependence of $A_0^{(0)}$  is almost flat due to the fact that $A_0(G)\approx -G$ 
but for increasing values of $G$ the presence of the bound state pole in
figure \ref{A_G} makes the variation range increasingly large.

The inclusion of first and second order corrections in (\ref{Luscher_A00}) -- i.e. $A_0^{(1)}$ and $A_0^{(2)}$  --
does not significantly improve this situation, specially when dealing
with large scattering length values.
This will be illustrated below.
On the other hand it is inconvenient to include higher order terms 
since the additional coefficients involved 
depend on dynamical parameters others than $A_0$ \cite{ML_CMP104_86}.

\begin{figure}[h!]
\begin{center}\mbox{\epsfxsize=8.cm\epsffile{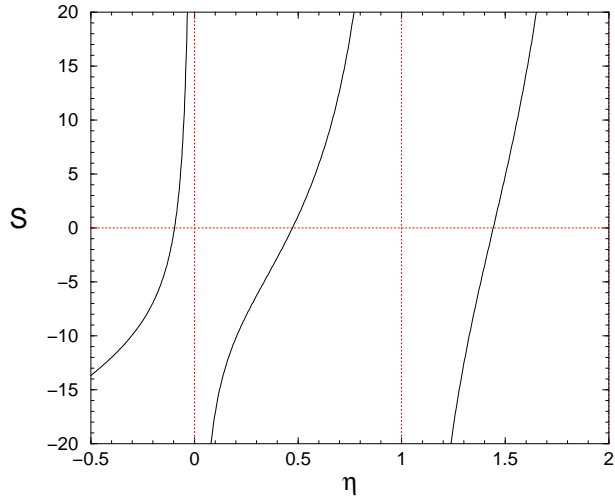}}\end{center}
\caption{Function $S(\eta)$ defined by (\ref{S_eta})}\label{Fig_S}
\end{figure}

Equation (\ref{Luscher_A}) is in fact a large-$L$ expansion of a more general relation 
established in \cite{ML_NPB354_91} between the two-body
phase shifts and the corresponding energy eigenstates on a finite size box.
For the ground state and using the notations of reference \cite{BBPS_PLB585_04}  it reads
\begin{equation}\label{Luscher_S}
k\cot\delta_0(k) = {1\over \pi aL}\; S(\eta)    
\end{equation}
where
\[\eta= {(aL)^2\epsilon_0(L)\over4\pi^2} = {A_0^{(0)}\over\pi aL} \]
$k^2=\epsilon_0(L)$ and $S$ is a universal function, independent of the interaction model,  defined by
\begin{equation}\label{S_eta}
S(\eta)=\lim_{\Lambda\to\infty}
 \sum_{n\in Z^3}^{|n|\le \Lambda} {1\over n^2-\eta}  -4\pi\Lambda
 \end{equation}
It is represented in figure \ref{Fig_S}. The domain of interest for the
ground state is $\eta<0$.

It is worth noticing that equation (\ref{Luscher_S}) 
is  exact  for lattice sizes $aL>2R$ where $R$ is the interaction range, i.e. $V(x>R)=0$.
For interactions of physical interest this regime
is reached exponentially and independently of the coupling constant and the $A_0$ value.
This constitutes a remarkable advantage with respect to expansion (\ref{Luscher_A}).

Using the effective range expansion (\ref{ER}) one finds the following relation
between the low energy parameters and the eigenenergies $\epsilon_0(L)$
\begin{eqnarray}\label{A0R0S}
-{1\over A_0}+{1\over2} 
R_0\epsilon_0(L)&=&{1\over\pi aL}S\left({(aL)^2\epsilon_0(L)\over4\pi^2}\right)
\end{eqnarray} 
By setting $R_0=0$, this relation can be written in the form
\[ \epsilon_0(L)= \left({2\pi\over aL}\right)^2 S^{-1}\left(-\pi {aL\over A_0}\right) \] 
where $S^{-1}$ denotes the reciprocal function of $S$. 
The Luscher expansion (\ref{Luscher_A}) 
is recovered by developping $S^{-1}$ for $z\to\infty$ in power series of $1/z$:
$S^{-1}(z)=-{1\over z} + {\pi c_1\over z^2} -{\pi^2c_2\over z^3} + \ldots $.

Equation (\ref{A0R0S}) suggests the possibility to obtain $A_0$ as a function of $\epsilon_0(L)$ \ldots provided
$R_0$ is known, which is in general never the case.

To check the applicability of (\ref{A0R0S})  in the Yukawa model we
fix the $R_0$ value from the infinite volume results
given on table \ref{Tab_LEP} and study the convergence of 
the scattering length thus obtained as  a function of the lattice size $aL$. 
We denote by $A^{(S)}_0(L)$ the quantity extracted this way, i.e.:
\begin{equation}\label{A0S}
A^{(S)}_0(L)  = \frac{\pi aL}{{1\over2}\pi\;R_0\;(aL)\epsilon_0(L)-S\left[ {(aL)^2\epsilon_0(L)\over4\pi^2}\right] }   
\end{equation}
It generalizes the series of values $A_0^{(n)}(L)$ define above.

\begin{figure}[h]
\begin{center}
\mbox{\epsfxsize=12.cm\epsffile{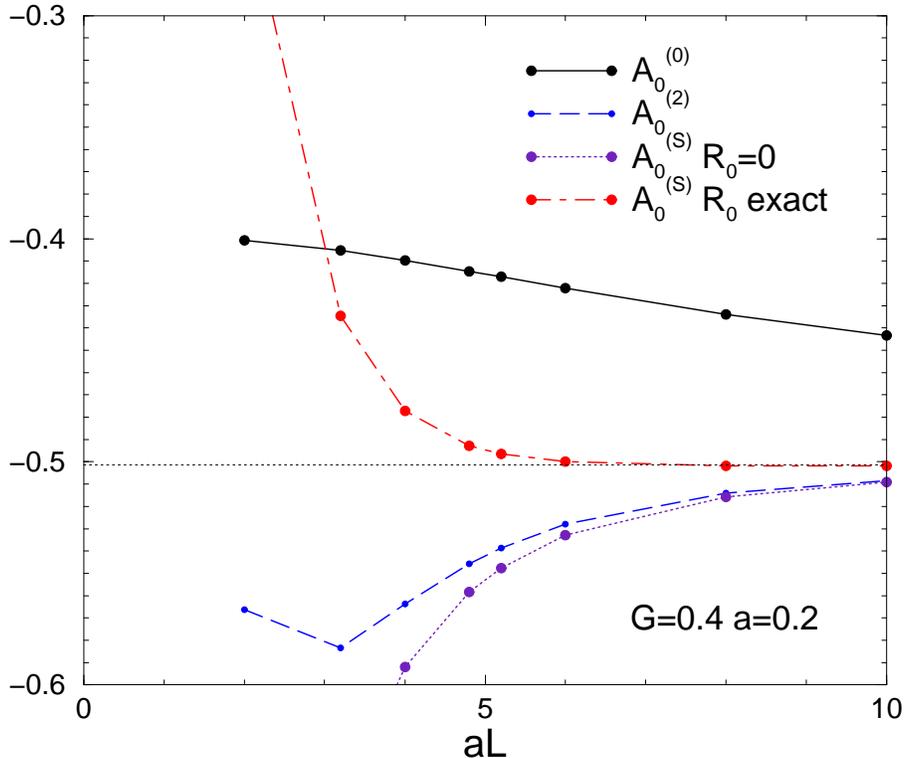}}
\caption{Different approximations used in calculating the
scattering lenght from energy eigenstates: $A_0^{(0)}$ 
defined in (\ref{A00}),  $A_0^{(2)}$ obtained by solving the cubic equation
(\ref{Luscher_A03}), $A_0^{(S)}$ given by (\ref{A0S}) with $R_0=0$ (dashed line)
and  with $R_0$ taken from the infinite volume results on table \ref{Tab_LEP} (dotted line).}\label{A00L_G=-0.4_a=0.20}
\end{center}
\end{figure}
Figure \ref{A00L_G=-0.4_a=0.20} shows
the results for $G=0.40$ and $a=0.20$, the same parameters than those used in figure
\ref{A00L_G=-0.4_LAMBDA}.
Only results including the fully lattice potential $V_L$ are hereafter shown.
Several remarks are in order: 
{\it (i)} $A^{(2)}_0$ represents an improvement
with respect to $A^{(0)}_0$ for $aL\gtrsim 5$, but it is significant
only in this particular case due to the smallness of ${A_0/aL}$ at small values of $G$
{\it (ii)}  $A^{(S)}_0$ is practically converged at $aL\approx 5$,
distance at which $A^{(0)}_0$ remains almost unchanged with respect to its value at $aL=0$
i.e. to the Born approximation
{\it (iii)}  the effective range contribution is sizeable:
if $R_0$  is neglected -- dashed curve obtained setting $R_0=0$ in (\ref{A0R0S}) --
the improvement with respect to $A^{(2)}_0$  -- and for $aL\lesssim5$ even to $A^{(0)}_0$  -- disappears.

\begin{figure}[h]
\begin{center}
\mbox{\epsfxsize=12.cm\epsffile{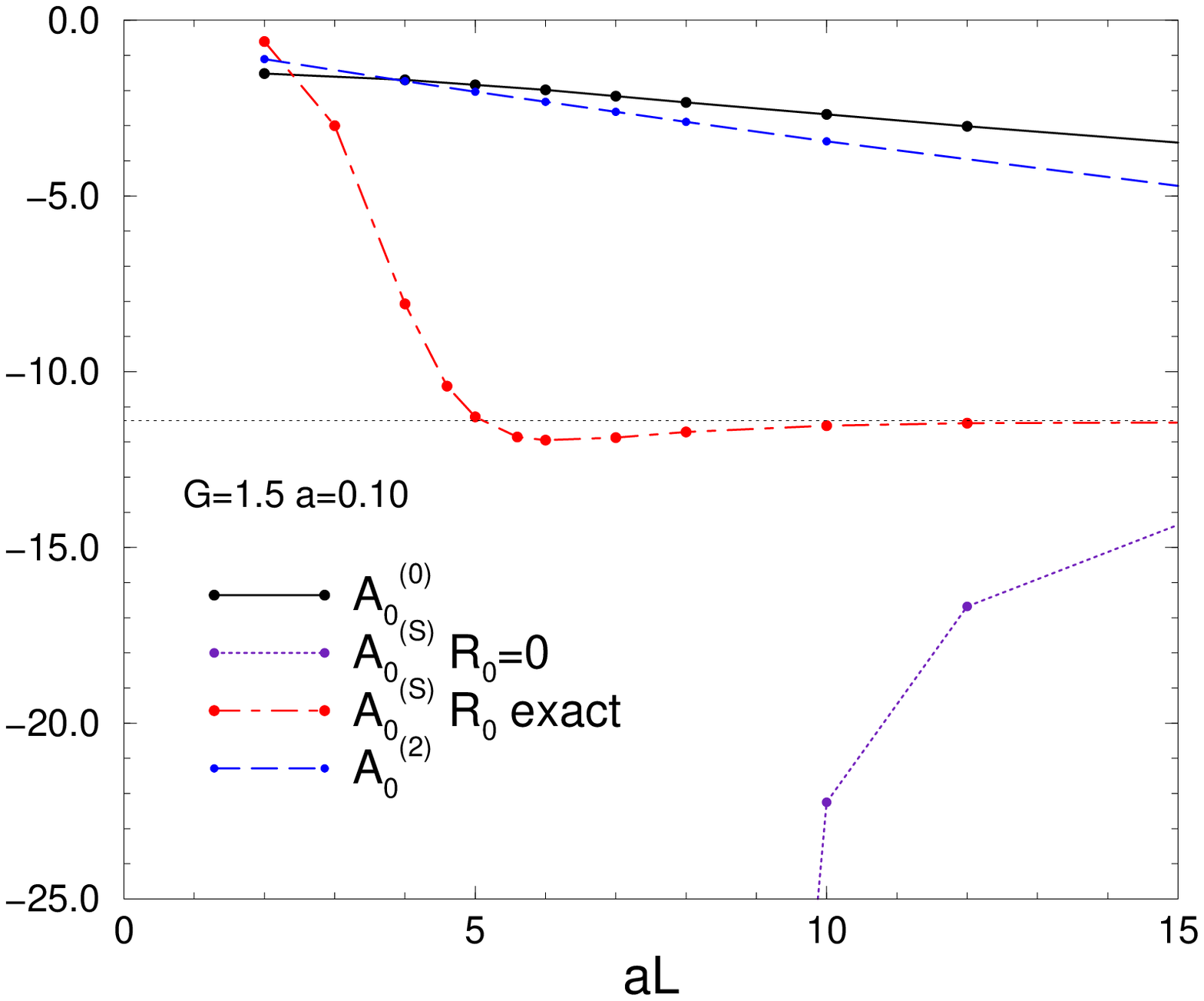}}
\caption{The same as figure (\ref{A00L_G=-0.4_a=0.20}) for a coupling constant $G=1.5$ near 
the bound state pole.
$A^{(S)}_0$  is practically converged at $aL\sim 5$, a lattice  size smaller than the infinite
volume scattering length $A_0(G)=11.4$ (horizontal dotted line).
Note that for small lattice, $A^{(0)}_0$ is practically constant -- i.e. $\epsilon_0(L)$ 
displays a  $1/L^3$ behaviour -- but this constant differs from $A_0$ by one order of magnitude .}
\label{A00L_G=-1.5_a=0.10}
\end{center}
\end{figure}

\bigskip
Similar results are displayed in figure \ref{A00L_G=-1.5_a=0.10}
for a value of the coupling constant $G=1.5$  near the bound state singularity
and a lattice space $a=0.10$. 
The infinite volume scattering length -- with $a=0.10$ -- is $A_0=-11.4$.
It corresponds to $a_0={A_0\over\mu}\approx -16$ fm if
a pion exchange ($\mu=0.14$ GeV) is supposed to govern the two-body interaction.
This value is close to the experimental $^1S_0$ nucleon-nucleon scattering length. 
One can see from this figure that, despite the large $A_0$ value, 
the convergence of $A^{(S)}_0(L)$ is the same as 
in Figure \ref{A00L_G=-0.4_a=0.20}: a few \% accuracy is already reached at $aL\approx 5$. 
The effective range contribution is here essential: by setting $R_0=0$ in (\ref{A0S}) 
one would get a result wrong by a factor two (dashed curve) even at $aL=10$.
Note that for small lattices, $A^{(0)}_0(L)$ is practically constant -- i.e. $\epsilon_0(L)$ 
displays a  $1/L^3$ behaviour -- but this constant differs from $A_0$ by one order of magnitude.
One can also remark the slow convergence of $A^{(2)}_0$, which, contrary to the  $G=0.4$ case, 
provides practically the same
results than $A^{(0)}_0$: even at $aL=10$ they differ from the asymptotic value $A_0$  by a factor 4.

\bigskip
The preceding results show the applicability of (\ref{A0R0S}) in extracting the 
scattering length $A_0$ for all cases of physical interest, provided that the effective range
$R_0$ is taken into account.
In practice, both values could be determined
by fitting the computed finite size energies $\epsilon_0(L)$ 
with the two-parameter function   (\ref{A0R0S}) for $aL\gtrsim 5$.
This procedure is however not very comfortable for the fitting
function (\ref{A0R0S}) is defined implicitly.
We alternatively propose to determine  $A_0$ and $R_0$ by
computing the energies $\epsilon_0(L)$ at two different lattice sizes near $aL=5$ and solving
the linear system resulting from (\ref{A0R0S}). 
This gives,  e.g. for the effective range
\begin{equation}\label{R0L1L2}
R_0={2\over\pi}
\frac{ {1\over aL_1}S\left({(aL_1)^2\epsilon_0(L_1)\over4\pi^2}\right)
-      {1\over aL_2}S\left({(aL_2)^2\epsilon_0(L_2)\over4\pi^2}\right) }
{\epsilon_0(L_1)-\epsilon_0(L_2)}
\end{equation} 

The results obtained using this procedure for $G=1.5$  are given in table \ref{Tab_A0R0S_G=1.5}. 
The corresponding infinite volume values calculated with $a=0.10$ are respectively $A_0=-11.4$ and $R_0=2.47$.
\begin{table}[h]
\begin{center}
\begin{tabular}{ r  r   c c}
$aL_1$& $aL_2$ & $R_0$& $A_0$     	\\\hline
4.0   &   4.60 & 1.9  & -28.8  \\
4.6   &   5.00 & 2.1  & -17.2  \\
5.0   &   5.60 & 2.3  & -13.9  \\
5.6   &   6.00 & 2.4  & -12.5  \\
6.0   &7.0     & 2.5  & -11.7   \\
7.0   &8.0     & 2.6  & -11.3   \\
8.0   &10.0    & 2.6  & -11.3   \\
10.0  &12.0    & 2.6 & -11.4   \\\hline
\multicolumn{2}{c}{Inf. volume} & 2.47 & 11.4 \\
\end{tabular}
\caption{Low energy parameters extracted from ground state energy at different lattice
sizes $\epsilon_0(L)$ by mean of equations (\ref{R0L1L2}) and (\ref{A0R0S}).
They correspond to $G=1.5$ and $a=0.1$.
Last row indicates the infinite volume results.}\label{Tab_A0R0S_G=1.5}
\end{center}
\end{table}
One sees from these results
that a lattice size $aL\approx6$ provides an $A_0$ value better than 3\% though the
accuracy for $R_0$ is slightly worse. 
The inclusion of an additional term -- $Pk^4$ -- in the low energy epansion (\ref{ER})
does not improve the results.

\begin{figure}[h]
\begin{center}
\mbox{\epsfxsize=12.cm\epsffile{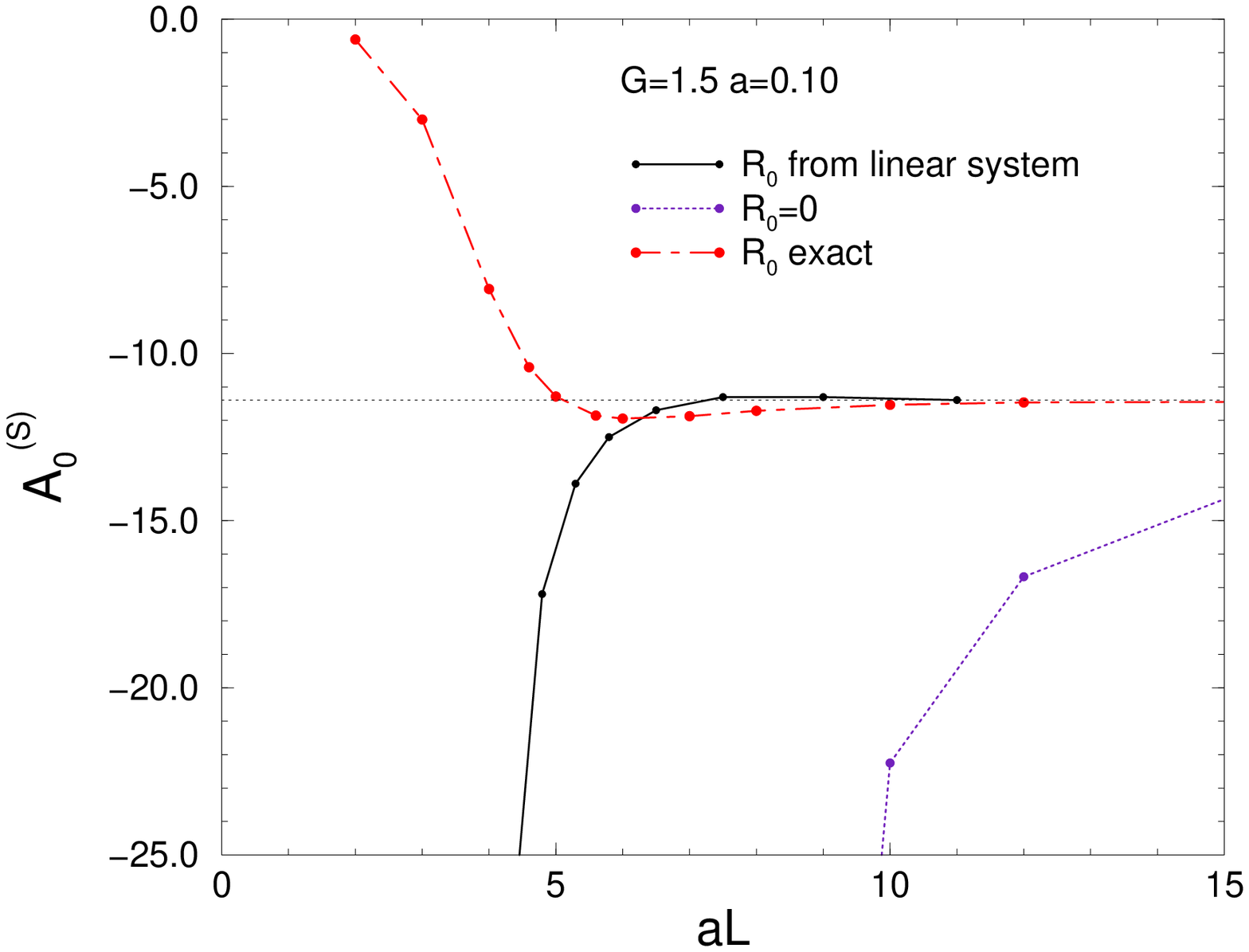}}
\caption{Different results obtained with equation (\ref{A0R0S})
depending on the value given to $R_0$.
Solid line correponds to the results of table \ref{Tab_A0R0S_G=1.5}
with an averaged value on abscissa $aL=0.5a(L_1+L_2)$ where
the two parameters are simultaneously determined. Dotted-dashed and dotted
correspond respectively to $R_0$ taken from Table \ref{Tab_LEP} and $R_0=0$.}
\label{A00L_G=-1.5_a=0.10_FIT}
\end{center}
\end{figure}

It could have some interest to summarize the different results obtained with equation (\ref{A0R0S})
depending on the way $R_0$ is taken into account.
This is done in figure \ref{A00L_G=-1.5_a=0.10_FIT} for $G=1.5$.
Solid line corresponds to the result of table \ref{Tab_A0R0S_G=1.5}
-- with an averaged value on abscissa $aL=0.5a(L_1+L_2)$ -- where
the two parameters $A_0$ and $R_0$ are simultaneously determined. Dotted-dashed and dotted
lines are the same than in figure  \ref{A00L_G=-1.5_a=0.10} 
and correspond respectively to $R_0$ taken from Table \ref{Tab_LEP}
and $R_0=0$.
Notice the different kind of convergence for $A_0^{(S)}$ 
and the key role of $R_0$.

\bigskip
This method applies well in all the range of coupling constants.
We would like to make some final comments concerning the extraction of $A_0$. 
The usual way of doing so is by first fitting the $L$-dependence
of the finite volume energies
by a $c/(aL)^3$ curve and identifying the coefficient $c$ with $4\pi A_0$.
In our notations, this is equivalent to use the $A^{(0)}_0$ value,
which is the leading term of  Luscher expansion (\ref{Luscher_A00}). 
As we have previously shown,
this approach can lead to wrong conclusions when using nowadays available lattice sizes,
unless the coupling constant -- and consequently the scattering length -- is very small.
In this case $A_0$ is already given by the Born approximation.
For the Yukawa model -- and for small values of the lattice size -- it actually
coincides with $-G$.

As one can see from figures \ref{A00L_G=-0.4_a=0.20} and \ref{A00L_G=-1.5_a=0.10}
$A^{(0)}_0(L)$ varies very smoothly even for almost resonant systems.
Thus, when using lattice sizes $aL\sim2-3$, $\epsilon_0(L)$ 
will be well fitted by a $c/L^3$ dependence but, as we have shown in these examples, 
the coefficient $c$ can strongly differ from the infinite volume scattering length.
Only the use of equation (\ref{A0R0S}) at lattice sizes greater
than the interaction range -- $aL\gtrsim5$ in the Yukawa model -- could lead to unambiguous extraction of the low
energy parameters, provided both $A_0$ and $R_0$ are taken into account.

Finally, we would like to notice that equation
(\ref{A000}) -- corresponding  to  a $\epsilon_0(L)\approx-{G\over (aL)^3}$ behaviour of 
the ground state eigenenergy -- depends only on general properties of the interaction 
as we have shown in Appendix \ref{Ap2}.
This result is totally independent of the Luscher relation (\ref{Luscher_S}) 
and they have even differents domains of applicability, namely $aL<<1$ and $aL>>1$. 
It is striking to note that $\epsilon_0(L)$ has, in both limits, the same $L$-dependence.

\begin{figure}[h!]
\begin{center}
\mbox{\epsfxsize=8.cm\epsffile{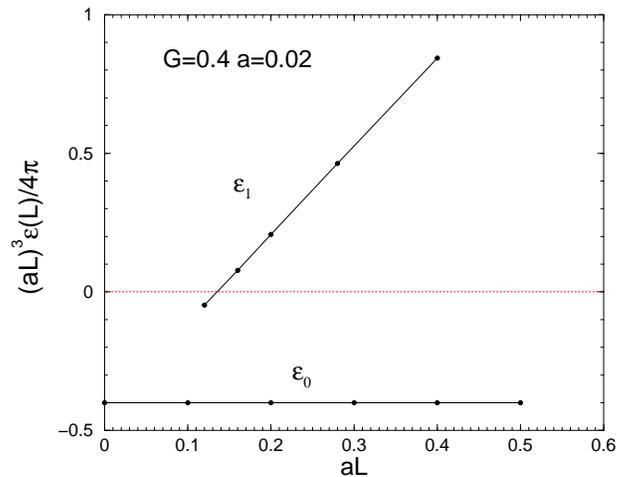}}
\caption{L-dependence of the ground $\epsilon_0$ and first energy $\epsilon_1$
states for $G=0.4$ in the limit of small lattice sizes.
All states converge to the same value, with
different slopes depending on their quantum number.}\label{A00L_G=-0.4_a=0.20_1st_Z}
\end{center}
\end{figure}

\bigskip
The results presented above concern the S-wave ground state energy $\epsilon_0(L)$.
For excited states, Luscher expansion (\ref{Luscher_A}) reads
\[  \epsilon_n(L)= \epsilon_n^{(0)}(L) +  \Delta_n(L)\]
with  $\epsilon_n^{(0)}(L)$ given by (\ref{eps0L})
and $\Delta_n(L)$ having the same form with differents values for the coefficients $c_i$.
In practice it has the same drawbacks than those described for the ground state.
Equation (\ref{A0R0S}) can be used, in principle, to obtain $A_0$ from the first excited state 
energy $\epsilon_1(L)$, now with  $\eta>0$.
At large values of $aL$, the L-dependence of excited states $\epsilon_n(L)$
is however dominated by the $1/(aL)^2$ term of
free solutions $\epsilon_n^{(0)}(L)$, thus making the calculations numerically more difficult.
It is worth noticing that in the limit $aL\to0$, all the excited states
tend to the same value than $\epsilon_0$ but this limit
is reached at lattice sizes increasingly small (see Appendix \ref{Ap2}).
This behaviour
is illustrated in figure \ref{A00L_G=-0.4_a=0.20_1st_Z} for the case $G=0.4$.

\subsection{Bound states}

\newcommand\eqtto{\mathop{=}}
Bound states are not affected by the "no go"
theorem in the euclidean formulation \cite{MT_NPB_245_90} 
and can by directly computed on lattice simulations at finite volume. 
However, the finite size effects are often non negligible in practice
and can  be controlled by similar expressions than those used for scattering states.
In addition, the determination of their binding energies in finite size lattice calculations
is made difficult by the existence of another length scale, independent
of the interaction range, given by the size of the state.
 
In his pioneer work \cite{ML_CMP104_86} Luscher established the following relation between the ground state 
energy $\epsilon_0$ of a non relativistic two body system and its value $\epsilon(L)$ 
computed in a finite size box with periodic boundary conditions
\cite{Rem2}:
\begin{equation}\label{Luscher_BS}
\epsilon(L)= \epsilon_0 -6 N_s^2\;{e^{-\kappa_0 aL}\over aL}  
\end{equation}
where $\kappa_0=\sqrt{-\epsilon_0}$ is the corresponding wave vector 
and $N_s$ the  asymptotic norm, defined in (\ref{N_s}), and determining the large $x$ behaviour of the wavefunction (\ref{Psi})

The two-body energy in a box $\epsilon(L)$
tends exponentially towards the infinite volume bound state binding energy $\epsilon_0$.
However its decreasing rate $\kappa_0$ contains  already the required information 
and can be consequently extracted well before the asymptotic region is reached.

\begin{figure}[h!]
\begin{minipage}[h!]{8.cm}
\begin{center}
\includegraphics[width=8.cm]{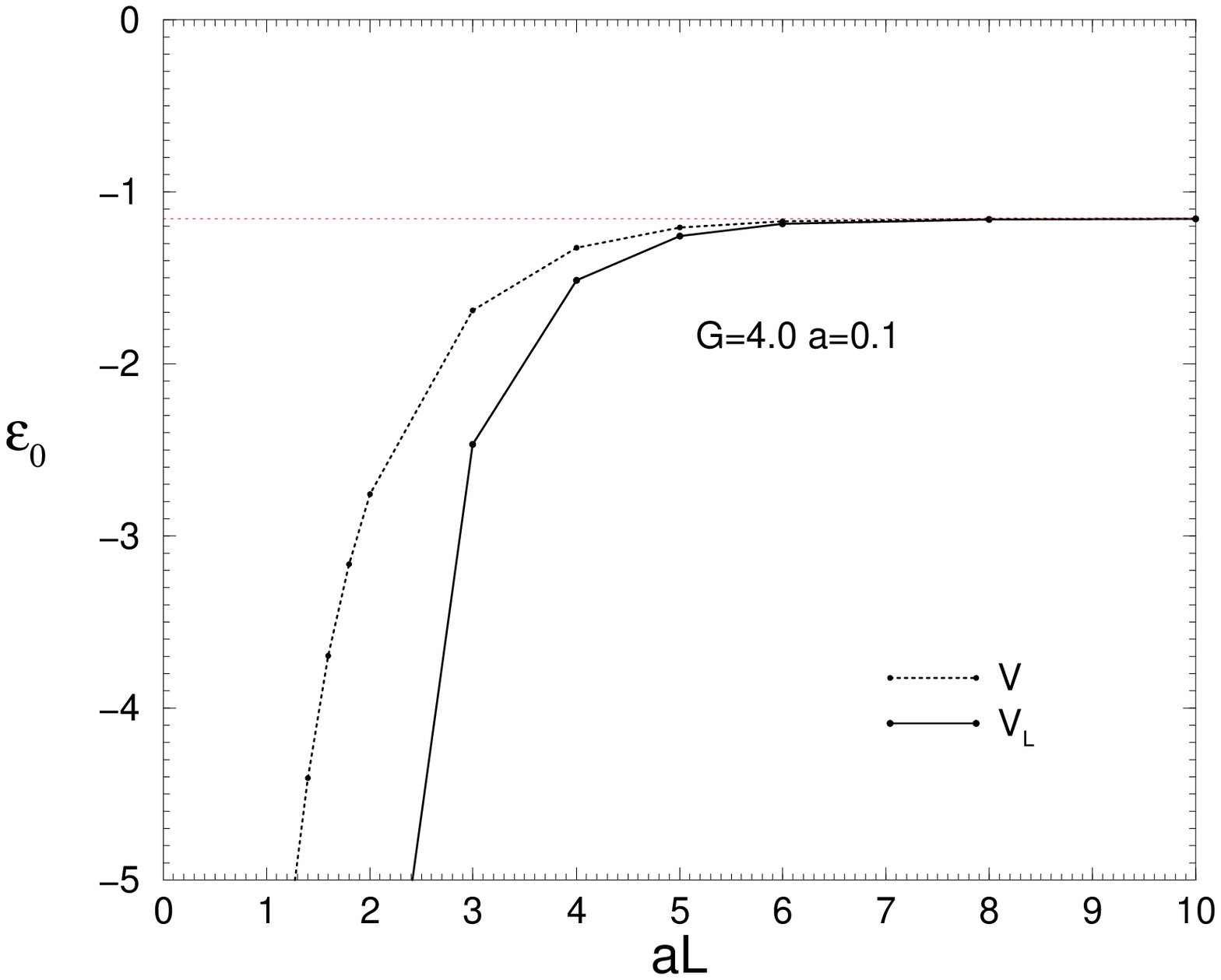}\end{center}
\caption{$L$-dependence of the ground state binding energy in the dimensionless Yukawa model
with $G=4$ and $a=0.1$,
Dashed line corresponds to one single
term $V$ in expansion (\ref{VL}) and solid line to the full interaction $V_L$.
Horizontal dotted line indicates the infinite volume result.}\label{EPSL_G=4.0}
\end{minipage}
\hspace{0.5cm}
\begin{minipage}[h!]{8cm}
\begin{center}\mbox{\epsfxsize=8.cm\epsffile{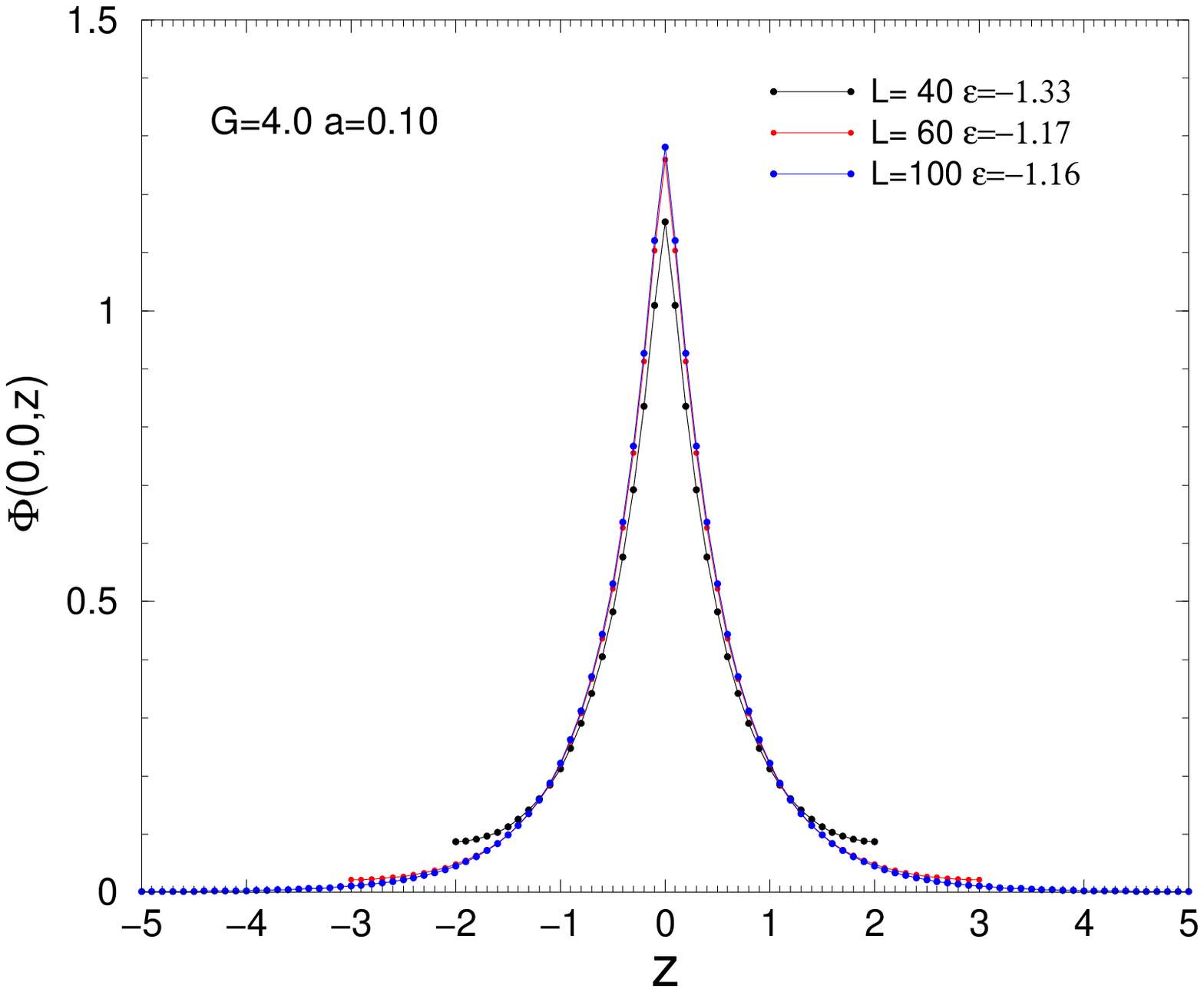}}\end{center}
\caption{Wavefunctions corresponding to figure \ref{EPSL_G=4.0} for different values of L}
\label{Psi_G=0.4}
\end{minipage}
\end{figure}

\bigskip
We first present the results concerning a deeply bound state. They have been
obtained with $G=4.0$ and $a=0.1$ and the $\epsilon(L)$ dependency
is displayed in Figure \ref{EPSL_G=4.0}. 
The asymptotic value $\epsilon_0=-1.16$ is  in good agreement
with the results of table \ref{Tab_BS} with the same lattice spacing $a=0.10$.
Although this value is reached -- at 1\% accuracy -- from $aL=6$, 
the same accuracy can be obtained by doing a two-parameter fit with function (\ref{Luscher_BS}) 
in the region $aL\in[3,4]$, where $\epsilon_0(L)$ is far from being converged.

Corresponding wavefunctions,  normalized according to (\ref{NORM}), are displayed in figure \ref{Psi_G=0.4} for different lattice sizes.
The periodic boundary conditions give rise to non vanishing
values at the lattice edges, which are reminiscent of the free ground state solution (\ref{FGSWF}).
The characteristic length of this state is $\rho=\sqrt{<x^2>}=1.07$, smaller
than the interaction range $R$ of the potential (\ref{Vad}).
One can thus expect that, like for the scattering states, its properties
would be well defined at $aL\sim 5$.
At distances greater than the interaction range, the wavefunctions
are not sensible to the boundary conditions and shows the typical
exponential behaviour of the bound states.
On the contrary at small values of lattice sizes, they look very similar
to the scattering states displayed in figure \ref{Phix_G=-0.4_a=0.20}.
The difference between bound and  scattering wavefunctions 
can be formalised by the parameter $s(L)={\Phi(L)\over\Phi(0)}$
which tends exponentially to 0 (bound) or like $1/(aL)$ to 1 (scattering) at large $aL$ values.

\bigskip
In analogy with what we have done for scattering states, it is possible
to extract both $\epsilon_0$ and $N_a$ by writing relation 
(\ref{Luscher_BS}) at two different lattice spacings.
The binding energy is thus determined by the solutions of the non linear equation
\begin{equation}\label{B_L1L2}
\frac{\epsilon_{L_1}-\epsilon_0}{\epsilon_{L_2}-\epsilon_0}=
\frac{L_2}{L_1}\; e^{-\sqrt{-\epsilon_0}\;a(L_1-L_2)}   
\end{equation}
Corresponding results for $G=4.0$ and $a=0.1$ are given in Table \ref{Tab_B_G=4.0}. 

\begin{table}[h]
\begin{center}
\begin{tabular}{ r  r  c c}
$aL_1$& $aL_2$ & $\kappa_0$& $\epsilon_0$     	\\\hline
2.0   &   3.0  & 1.14 & -1.30  \\
3.0   &   4.0  & 1.09 & -1.19  \\
4.0   &   5.0  & 1.08 & -1.16  \\
5.0   &   6.0  & 1.08 & -1.16  \\\hline
\multicolumn{2}{c}{Inf. volume} & 1.07  & -1.15 \\
\end{tabular}
\caption{Ground states energy for $G=4.0$ and $a=0.1$
by solving equations (\ref{B_L1L2}) at different lattice sizes.
Infinite volume results are taken from table \ref{Tab_BS}
with the same value of $a$.}\label{Tab_B_G=4.0}
\end{center}
\end{table}

The binding energy and characteristic length of this example 
roughly correspond to the physical case of $^4$He nuclei. 
Indeed, by inserting the $\epsilon=-1.16$ and ${\mu\over M}=0.15$
in equation (\ref{E}) one gets a binding energy in units of nucleon mass $E/M=-0.026$
while the experimental value for $^4He$ is $E_{^4He}/M=-0.030$.

\begin{figure}[h!]
\begin{minipage}[h!]{8.cm}
\begin{center}\includegraphics[width=8.cm]{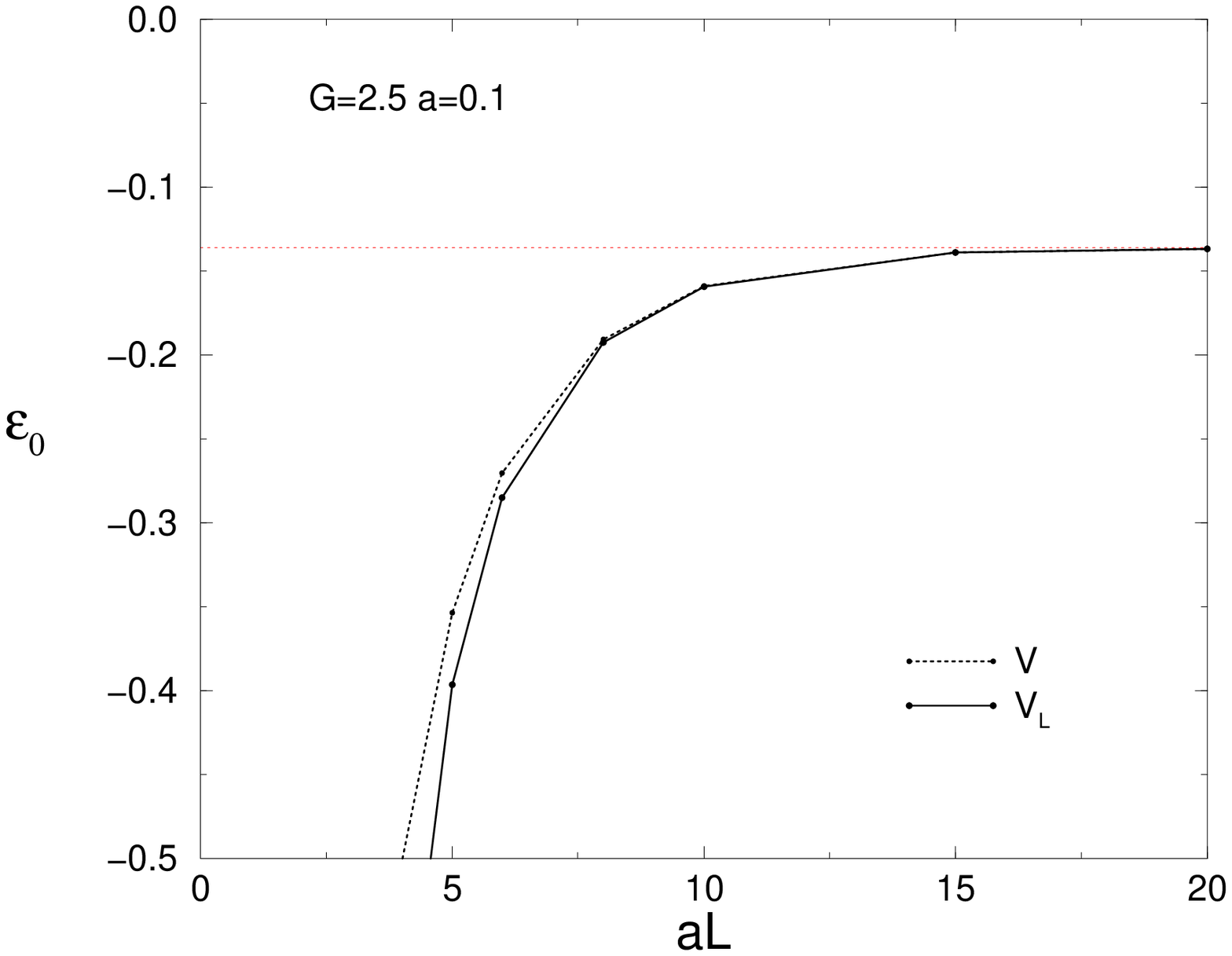}\end{center}
\caption{$L$-dependence of the ground state energy 
for $G=2.5$ and $a=0.1$. Notations are the same than for figure \ref{EPSL_G=4.0}}\label{EPSL_G=2.5}
\end{minipage}
\hspace{0.5cm}
\begin{minipage}[h!]{8cm}
\begin{center}\mbox{\epsfxsize=8.cm\epsffile{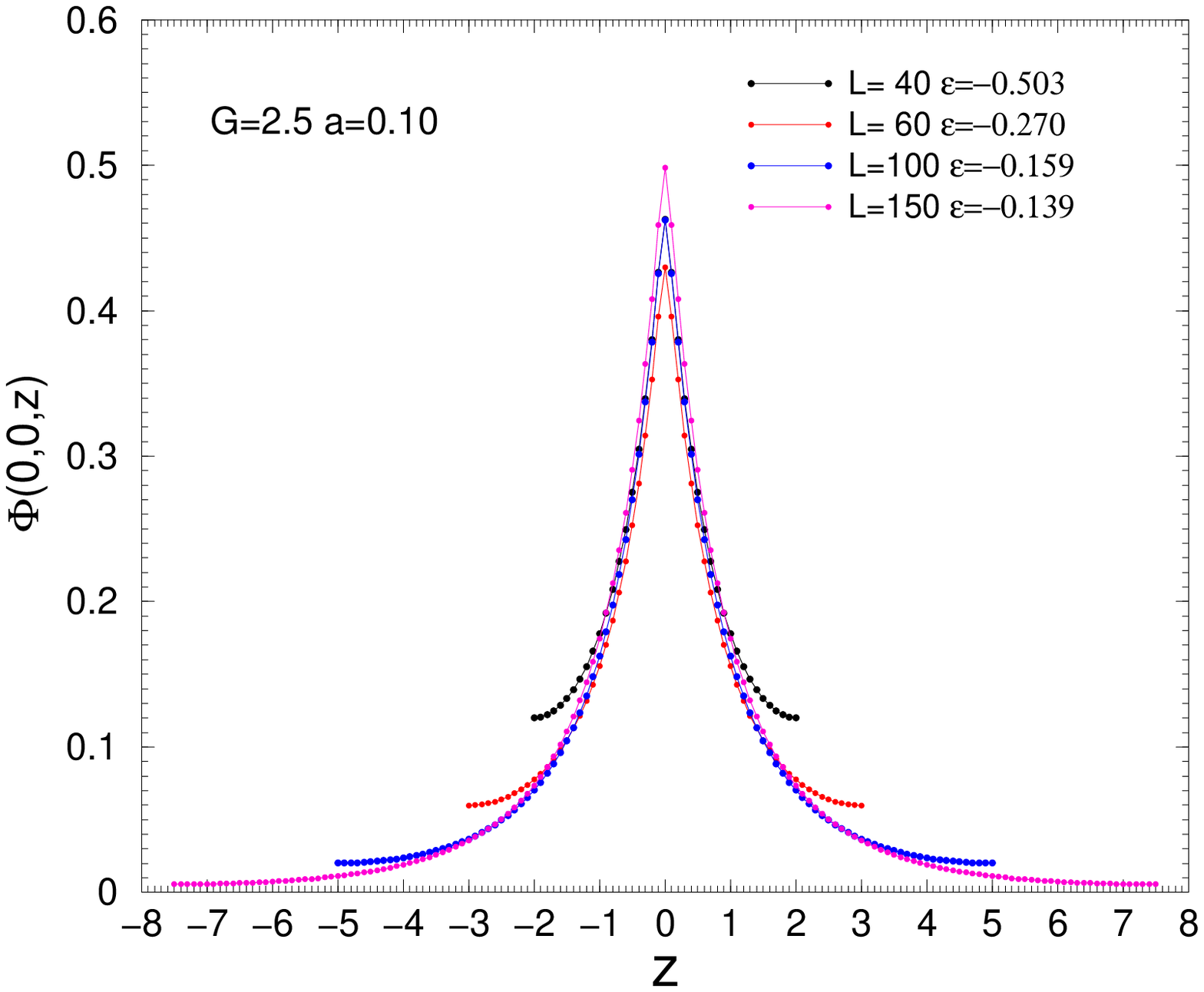}}\end{center}
\caption{Wavefunctions corresponding
to figure \ref{EPSL_G=2.5} for different values of L}\label{Psi_G=2.5}
\end{minipage}
\end{figure}

\bigskip
The situation is less comfortable when dealing with loosely bound states.
An example, close to the deuteron, is obtained with $G=2.5$
for which $\epsilon=-0.136$ and, according to (\ref{E}), $E/M=-0.003$.
Corresponding results are displayed in figures \ref{EPSL_G=2.5} and \ref{Psi_G=2.5}.
The convergence of $\epsilon(L)$ is reached in this case at $aL\geq15$ due
to the larger size of the state, for which $\rho=2.5$.
It is however possible, by using equation (\ref{B_L1L2}), to reach already a few percent accuracy
in the range $aL\in[8,10]$. Results obtained this way are given in Table \ref{Tab_B_G=2.5}.

\begin{table}[h]
\begin{center}
\begin{tabular}{ r  r   cc}
$aL_1$&$aL_2$& $\epsilon_0$& $N_s$  	 \\\hline
5.0   & 6.0  &-0.131	    &  1.2\\
6.0   & 8.0  &-0.143	    &  1.2\\
8.0   &10.0  &-0.139	    &  1.2\\
10.0  &15.0  &-0.137	    &  1.2\\
15.0  &20.0  &-0.137	    &  1.2\\\hline
\multicolumn{2}{c}{Inf. volume}&-0.136        &  1.17    \\ 
\end{tabular}
\caption{Ground states energy $\epsilon_0$ and asymptotic norm $N_s$ 
for $G=2.5$ and $a=0.1$ 
obtained by solving equation (\ref{B_L1L2}) at different lattice sizes.}\label{Tab_B_G=2.5}
\end{center}
\end{table}

These examples indicate that the critical lattice size is not given by the 
interaction range but by the size of the bound state itself.
It is nevertheless possible by means of equation (\ref{B_L1L2}) to access the 
binding energies of loosely bound states like deuteron.
Notice however, that in order to have a few percent accuracy
on the binding energy, a consequent number of lattice points $L\sim 100$ 
would be in this case required.

\begin{figure}[h!]
\begin{center}\mbox{\epsfxsize=10.cm\epsffile{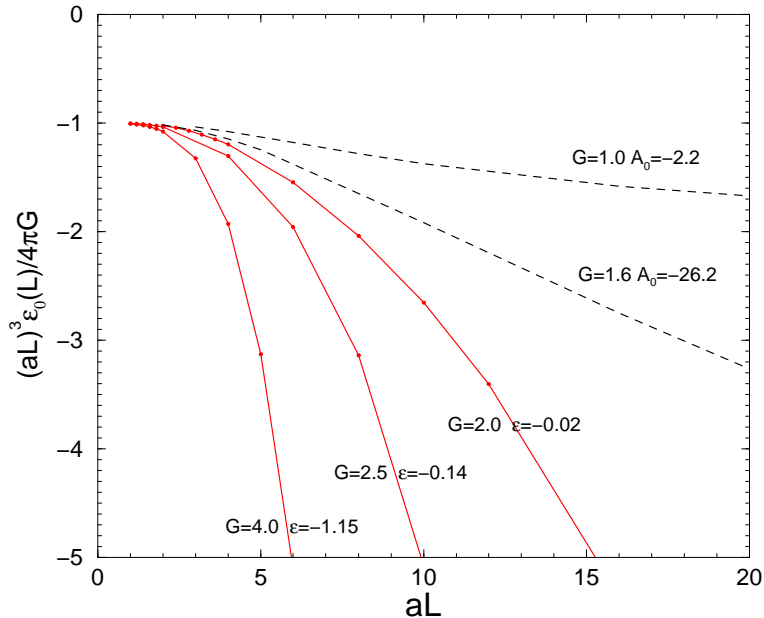}}\end{center}
\caption{Comparison of the $(aL)^3\epsilon_0(L)/(4\pi G)$ behaviour for
bound (solid lines) and scattering (dotted lines) states.
At $aL\to0$ they all tend to $-1$ as given by (\ref{A000}) but at large $aL$,
they manifest either a cubic divergence or a slow $1/L$ convergence
towards the asymptotic value $A_0(G)/G$.}\label{L3EPSL_ALLG_NORM_Z2}
\end{figure}

The limit $aL\to 0$ of $\epsilon_0(L)$ obtained in Appendix \ref{Ap2}, 
holds  for bound as well as for scattering ones: 
in both cases the eigenenergies display the $1/L^3$ law given by equation (\ref{A000}). 
The differences in $\epsilon_0(L)$ appear clearly in the regions
where the interaction plays no role, which in the Yukawa model is at $aL\approx 5$.
They are manifested by calculating the quantity $(aL)^3\epsilon_0(L)/(4\pi G)$ 
-- which corresponds to $A^{(0)}_0/G$ of the previous section -- 
and comparing the results for bound and scattering states.
This has been done in figure \ref{L3EPSL_ALLG_NORM_Z2} for increasing values of G.  
For the bound state case (solid lines) one can see  
a cubic divergence corresponding to $\epsilon_0$ term  in (\ref{Luscher_BS}).
For scattering states (dashed lines) one observe  
a linear dependence before reaching an horizontal asymptote $A_0(G)/G$.
Both behaviours are easy to disentangle in the extreme cases 
of deeply bound state or small scattering length but 
it becomes increasingly difficult when approaching resonant scattering and/or 
loosely bound state, unless huge number of lattice points are used \cite{SY_05}.

\section{Conclusion}\label{concl}

We have obtained the solutions of the non relativistic Yukawa model (\ref{V})
in a 3-dimensional lattice with periodic boundary conditions
and examined the possibility to extract infinite volume low energy parameters
--  scattering length $A_0$ and effective range $R_0$ --
and bound state binding energies from the computed eigenstates $\epsilon_n(L)$ at 
finite lattice sizes $aL$.
The eigenergies $\epsilon_n(L)$, and corresponding eigenfunctions, have been calculated 
using finite difference schemes and propagation in the euclidean time,
in close analogy with the methods used in lattice field theory simulations.
The low energy parameters 
have been independently computed  by solving the corresponding 
one-dimensional Schrodinger radial equation.

\bigskip
We have considered different approximations of the Luscher relations and shown
that a lattice size of $aL\approx 5$ -- in units of the meson exchanged Compton 
wave lenght ${1\over \mu}$ --
is enough to determine the low energy parameters at few \% accuracy.
The method is based in computing the energies at two different lattice
sizes and solving a linear system of rank two.
Such a possibility is independent of the scattering length value and applies
in the resonant case as well. 
We have found in particular that the effective range $R_0$ plays a crucial role in determining
the scattering length value at  moderate lattice volumes.

\bigskip
We have shown that in the limit of small lattice sizes, the 
L-dependence of eigenenergies is dominated by an $\epsilon_n(L)=-G/(aL)^3$ 
term, where $G$ is the strength parameter of the dimensionles potential (\ref{Vad}).
This regime is already reached at moderate sizes $aL\lesssim 2$, 
making ambiguous the extraction
of the scattering lenght by means of Luscher expansion (\ref{Luscher_A}).
The behaviour is the same for bound and scattering states
and applies to a large class of potentials. 

\bigskip
For loosely bound states, the critical
lattice size is determined by the spatial extension $\rho$ of the state.
We have shown that for the deuteron case
it is however possible to get accurate values of the
binding energy and asymptotic norm with $aL=10$.
The parameters of deeply bound states are well determined,
like for the scattering case, with lattice volumes $aL\approx 5$.

\bigskip
The results have been obtained with a non relativistic model,
which is justified by the small energies involved in the calculations.
Despite its simplicity, the model
considered  contains an essential ingredient of the hadron-hadron interaction
-- its finite range -- which plays a relevant role
in view of extracting the low energy parameters from the finite volume spectra.
It offers a wieldy and physically sound tool to test the validity of the different
approaches discussed in the literature to study the low energy scattering 
of  baryon-baryon or meson-baryon systems from a lattice simulations in QCD.

\bigskip
The results presented in this work have been essentially limited to the ground 
state of central attractive interactions, depending  only on one paremeter.
The method can be easily applied to more involved interactions, 
like hard core repulsive terms or non central potentials
leading to coupled channel equations.

\section*{Acknowledgement}
We are warmly grateful to our colleagues J.P. Leroy, Ph. Boucaud, O. Pene 
(L.P.Th. Orsay) and C. Roiesnel (CPhT Ecole Polytechnique)
for many useful discussions during the elaboration of this work.
We are indebted to S. Sint 
for pointing out relevant aspects of Luscher's work during his visit to LPSC. 
We thanks M. Mangin-Brinet for a critical reading of the manuscript.

\appendix
\section{Born approximation}\label{Ap1}

Let us consider the solution of (\ref{Sch_RR})
in terms of the equivalent Lipmann-Schwinger equation for the K-matrix
\[ K= U+ {\rm P.V.}\left\{  UG_0K\right\}\]
where P.V. denotes the principal value.
The lowest order in the coupling constant ($g^2$) is given by the Born term, i.e.
\[ K= U \]
The  K-matrix is related to the S-wave phaseshifts $\delta_0$ at momentum $k_0$
by \cite{GL_83}
\[  K_0(k_0,k_0)= -\frac{1}{(2\pi)^2m_R}\; \frac{\tan\delta_0(k_0)}{k_0} \]
where $K_0$ denotes the on-sell $L=0$ term of the $K$-partial wave 
expansion and $m_R$ the reduced mass of the system.
In the limit $k_0\to0$ on gets
\[ a_0 =(2\pi)^2 m_R K_0(0,0)\] 
The lowest order (Born approximation) is thus given by
	\[ a_0(g) =(2\pi)^2 m_R U(0,0) + o(g^4)\] 
where $U$ must be understood as the $\vec{k}=\vec{k}'=0$ matrix element
of  the potential in momentum space $U(\vec{k},\vec{k'})$.
For the Yukawa model one has
\[ U(\vec{k},\vec{k'})= -{1\over(2\pi)^3} \;{g^2\over q^2+\mu^2}  
\qquad \vec{q}=\vec{k}-\vec{k'}  \]
and in case of two identical particles with mass $M=2m_R$ 
\[ a_0(g) =- {g^2\over4\pi}\; {M\over\mu^2} +o(g^4) \] 
In dimensionless units (\ref{a_0}) it reads
\[ A_0(G)= -G +o(G^2) \]
which is our equation (\ref{A0G}).

\section{Small volume limit}\label{Ap2}

We would like to study here the zero volume limit of equation (\ref{Sch_3D}) 
with periodic boundary conditions and  a potential of the form (\ref{VL}).
To this aim, it is useful to develop  the periodic wave function in a Fourier series:
\[\Phi(x)=\sum_{\vec{n}\in Z^3} C_{\vec{n}}\; e^{\frac{2i\pi}{aL}\vec{n}\cdot \vec{x}}\ ,\]
and similarly for the potential,
\[V_L(x) = \sum_{\vec{n}\in Z^3} V_{\vec{n}}\; e^{\frac{2i\pi}{aL}\vec{n}\cdot \vec{x}}\ \]
By inserting these expresions into equation (\ref{Sch_3D}) 
it results  the following infinite system of coupled linear equations:
\begin{equation}\label{fourier}
\left[\left(\frac{2\pi}{aL}\right)^2\vec{n}^2-E\right]C_{\vec{n}}
+\sum_{\vec{n}'\in Z^3} V_{\vec{n}-\vec{n}'} C_{\vec{n}'}  = 0\ .
\end{equation}

The Fourier coefficients  of the potential, $V_{\vec{n}}$,
take a rather simple form due to the interactions with the "surrounding world",
\begin{eqnarray}
V_{\vec{m}} 
&=& \frac{1}{(aL)^3} \int_0^{aL} d^3{\vec{x}} \; e^{-{2i\pi\over aL}{\vec{m}}\cdot\vec{x}}
\sum_{\vec{n}}V(\vec{x}+aL\vec{n})\nonumber\\
&=& \frac{1}{(aL)^3}\sum_{\vec{n}}\int_0^{aL} dx_1 \int_0^{aL}
dx_2\int_0^{aL} dx_3\ e^{-{2i\pi\over aL}{\vec{m}}\cdot\vec{x}}\;V(\vec{x}+aL\vec{n})\nonumber\\
&=& \frac{1}{(aL)^3}\sum_{\vec{n}}\int_{n_1
aL}^{(n_1+1)aL}dy_1\int_{n_2 aL}^{(n_2+1)aL}dy_2\int_{n_3
aL}^{(n_3+1)aL}dy_3 \ e^{-{2i\pi\over aL}{\vec{m}}\cdot\vec{y}}\;V(\vec{y})\nonumber\\
&=& \frac{1}{(aL)^3} \int d^3\vec{y}\ e^{-\frac{2i\pi}{aL}\vec{m}\cdot\vec{y}}
\;V(\vec{y}) \label{Vm}
\end{eqnarray}
For the dimensionless  Yukawa potential (\ref{Vad}) they read:
\[
V_{\vec{m}} = \left\{ \begin{array}{l r} -\frac{4\pi G}{(aL)^3} &
\vec{m}=(0,0,0) \\ & \\
-\frac{4\pi G}{(aL)^3}\frac{\frac{m\pi}{aL}}{1+4\left(\frac{m\pi}{aL}\right)^2} & \vec{m}\neq(0,0,0) \\
\end{array} \right.
\]
The leading term in the small volume limit is given by $V_{\vec{0}}$. 
Neglecting other components, the system of equations (\ref{fourier}) decouple
into
\[
\left[\left(\frac{2\pi}{aL}\right)^2 n^2-\epsilon\right]C_{\vec{n}}+V_{\vec{0}}C_{\vec{n}}= 0\ ,
\]
and the eigenenergies are given by:
\begin{equation}
 \epsilon_n(L)= V_{\vec{0}} + \left(\frac{2\pi}{aL}\right)^2 n^2
= -\frac{4\pi G}{(aL)^3}+\left(\frac{2\pi}{aL}\right)^2 n^2\ 
\end{equation}
They correspond to the continuum limit of the free result (\ref{eps0L}) shifted by 
a constant potential of depth $V_{\vec{0}}$. 

One has for the ground state
\begin{equation}
\epsilon_0(L)=-\frac{4\pi G}{(aL)^3}
\end{equation}
a result which  is equivalent, and prove, equation (\ref{A000}).
Notice that in the limit $aL\to0$ all the excited states $\epsilon_n$
display also the same $1\over (aL)^3$ behaviour 
and furthermore they all tend to the same value $\epsilon_0(L)$, independently of $n$.
The lattice size at which they become negative $aL= {G\over \pi n^2}$
are however increasingly small.


\end{document}